\documentclass{jfm}

\usepackage{amsmath}
\usepackage{graphicx}
\usepackage{natbib}
\usepackage{colortbl}
\usepackage{multicol}
\usepackage[usenames,dvipsnames,svgnames,table]{xcolor}
\usepackage{stmaryrd}

\ifCUPmtlplainloaded \else
  \checkfont{eurm10}
  \iffontfound
    \IfFileExists{upmath.sty}
      {\typeout{^^JFound AMS Euler Roman fonts on the system,
                   using the 'upmath' package.^^J}%
       \usepackage{upmath}}
      {\typeout{^^JFound AMS Euler Roman fonts on the system, but you
                   dont seem to have the}%
       \typeout{'upmath' package installed. JFM.cls can take advantage
                 of these fonts,^^Jif you use 'upmath' package.^^J}%
       \providecommand\upi{\pi}%
      }
  \else
    \providecommand\upi{\pi}%
  \fi
\fi

\ifCUPmtlplainloaded \else
  \checkfont{msam10}
  \iffontfound
    \IfFileExists{amssymb.sty}
      {\typeout{^^JFound AMS Symbol fonts on the system, using the
                'amssymb' package.^^J}%
       \usepackage{amssymb}%

      }{}
  \fi
\fi

\ifCUPmtlplainloaded \else
  \IfFileExists{amsbsy.sty}
    {\typeout{^^JFound the 'amsbsy' package on the system, using it.^^J}%
     \usepackage{amsbsy}}
    {\providecommand\boldsymbol[1]{\mbox{\boldmath $##1$}}}
\fi

\newcommand{\units}[1]{\mbox{$\mathrm{#1}$}}

\newcommand\Real{\mbox{Re}} 
\newcommand\Imag{\mbox{Im}} 

\newsavebox{\astrutbox}
\sbox{\astrutbox}{\rule[-5pt]{0pt}{20pt}}

\newcommand{\iim}{\mathrm{i}}

\renewcommand{\vec}[1]{\boldsymbol{#1}}

\newcommand\bnabla{\boldsymbol{\nabla}}
\newcommand\bcdot{\boldsymbol{\cdot}}

\renewcommand{\u}{\boldsymbol{u}}
\newcommand{\f}{\boldsymbol{f}}
\newcommand{\F}{\boldsymbol{F}}
\newcommand{\x}{\boldsymbol{x}}
\newcommand{\X}{\boldsymbol{X}}

\newcommand{\mat}[1]{\mathsfbi{#1}}

\newcommand{\gameps}{\varrho}
\newcommand{\Oe}{{\Omega_\gameps(t)}}
\newcommand{\dOe}[1]{{\partial\Omega_\gameps^{#1}(t)}}

\newcommand{\unitR}{\hat{\boldsymbol{r}}}
\newcommand{\unitTheta}{\skew3\hat{\boldsymbol{\theta}}}
\newcommand{\unitPhi}{\skew3\hat{\boldsymbol{\phi}}}

\newcommand{\normal}{\hat{\boldsymbol{n}}}

\newcommand{\vshY}{\boldsymbol{Y}_{m,k}}
\newcommand{\vshPsi}{\boldsymbol{\Psi}_{m,k}}
\newcommand{\vshPhi}{\boldsymbol{\Phi}_{m,k}}

\newcommand{\iints}{\int_{0}^{\upi}\hspace{-5pt} \int_{0}^{2\upi}}

\renewcommand{\d}{\mathrm{d}}

\newcommand{\jump}[1]{\left[\hspace{-1pt}\left[#1\right]\hspace{-1pt}\right]}
\renewcommand{\jump}[1]{\llbracket#1\rrbracket}
\newcommand{\bigjump}[1]{\bigg\llbracket#1\bigg\rrbracket}

\newcommand{\half}{\frac{1}{2}}

\everymath{\displaystyle}

\title[Parametric resonance in spherical immersed elastic shells]%
{Parametric resonance \\ in spherical immersed elastic shells}

\author[W. Ko and J. M. Stockie]%
{William Ko$^1$\thanks{Email address for correspondence: wka11@sfu.ca}\ns
  and John M. Stockie$^1$}

\affiliation{$^1$Department of Mathematics, Simon Fraser University,
Burnaby, BC, V5A 1S6, Canada}

\pubyear{}
\volume{}
\pagerange{}
\date{?; revised ?; accepted ?. - To be entered by editorial office}
\begin{document}

\maketitle

\begin{abstract}
  We perform a stability analysis for a fluid-structure interaction
  problem in which a spherical elastic shell or membrane is immersed in
  a 3D viscous, incompressible fluid.  The shell is an idealised
  structure having zero thickness, and has the same fluid lying both
  inside and outside.  The problem is formulated mathematically using
  the immersed boundary framework in which Dirac delta functions are
  employed to capture the two-way interaction between fluid and immersed
  structure.  The elastic structure is driven parametrically via a
  time-periodic modulation of the elastic membrane stiffness.  We
  perform a Floquet stability analysis, considering the case of both a
  viscous and inviscid fluid, and demonstrate that the forced
  fluid-membrane system gives rise to parametric resonances in which the
  solution becomes unbounded even in the presence of viscosity.  The
  analytical results are validated using numerical simulations with a 3D
  immersed boundary code for a range of wavenumbers and physical
  parameter values.  Finally, potential applications to biological
  systems are discussed, with a particular focus on the human heart and
  investigating whether or not FSI-mediated instabilities could play a
  role in cardiac fluid dynamics.
\end{abstract}



\section{Introduction}
\label{sec:intro}
 
Fluid-structure interaction (or FSI) problems are ubiquitous in
scientific and industrial applications.  Because of the complex coupling
that occurs between the fluid and moving structure, FSI problems present
formidable challenges to both mathematical modellers and computational
scientists.  Of particular interest in this paper are flows involving a
highly deformable elastic structure immersed in a viscous fluid, which
is common in bio-fluid systems such as blood flow
in the heart or arteries, dynamics of swimming or flying organisms,
and organ systems \citep{Kleinstreuer2006, Lighthill1975, Vogel1994}.

One approach that has proven particularly effective at capturing FSI
with highly deformable structures is the immersed boundary (or IB)
method \citep{Peskin2002}.  This approach was initially developed by
\citet{Peskin1977} to study the flow of blood in a beating heart, and
has since been employed in a wide range of other biological and
industrial applications.  The primary advantage of the IB method is its
ability to capture the full two-way interaction between an elastic
structure and a surrounding fluid in a simple manner using Dirac delta
function source terms, which also leads to a simple and efficient
numerical implementation.

A common test problem that is often employed to validate both
mathematical models and numerical algorithms in FSI problems involves an
oscillating spherical elastic shell immersed in fluid, very much like a
rubber water balloon immersed in a water-filled container.  The
two-dimensional version of this problem is the standard circular
membrane problem that pervades the IB literature.  Extensive research
has also been performed on numerical simulations of immersed spherical
shells that aims to develop accurate and efficient algorithms, in
applications that range from red blood cell deformation in capillaries
\citep{Ramanujan1998}, to sound propagation in the cochlea
\citep{Givelberg2004}, to contraction of cell membranes
\citep{Cottet2006}.

Outside of numerical simulations, relatively little mathematical
analysis has been performed on such FSI problems owing to the complex
nonlinear coupling between the equations governing the elastic structure
and the fluid in which it is immersed.  \citet{Stockie1995} performed a
linear stability analysis of the IB model in two dimensions for a simple
geometry in which a flat membrane or sheet is immersed in fluid. They
also presented asymptotic results on the frequency and rate of decay of
membrane oscillations that depends on the wavenumber of the sinusoidal
perturbation, as well as validating their analytical results using 2D
numerical simulations with the IB method.  \citet{Cortez1997} performed
a non-linear analysis of a perturbed circular elastic membrane immersed
in an inviscid fluid.  These results were later extended by
\citet{Cortez2004} to the study of parametric instabilities of a
internally-forced elastic membrane in two dimensions, in which the
forcing appears as a periodic modulation of the elastic stiffness
parameter.  These authors showed that such systems can give rise to
parametric resonances, which overcome viscous fluid damping and thereby
cause the elastic structure to become unstable.  A similar analysis was
performed by \citet{KoStockie2014} on a simpler two-dimensional
flat-membrane geometry, and applied to the study of internally-forced
oscillations in the mammalian cochlea.  Parametric resonance is a
generic form of unstable response that can arise whenever there is a
time-periodic variation in a system parameter~\citep{Champneys2009}, and
such resonances have been identified in a wide range of fluid and
biofluid systems, for example by~\citet{Gunawan2005, Kelly1965,
  KumarTuckerman1994, SemlerPaidoussis1996}.

In this article, we extend the work of \citet{Cortez2004} to three
dimensions by studying parametric instabilities of an elastic spherical
shell immersed in a incompressible, Newtonian fluid.  The internal
forcing is induced by a time-periodic modulation of the membrane
stiffness parameter.  The most closely-related study in the literature
is a paper by \citet{Felderhof2014} that investigates the response of an
elastic shell to an impulsive acoustic forcing, which differs
significantly in that it is not only an externally forced problem but
also involves much more energetic excitations at much higher
frequencies.  Our work is motivated in part by IB models of active
biological systems such as the heart, wherein the contraction and
relaxation of heart muscles can be mimicked by a time-dependent muscle
stiffness.  Furthermore, \citet{Cottet2006a} presented computational evidence for
the existence of parametric instabilities, but no analysis has yet been
done of the complete three-dimensional governing equations to confirm
the presence of these instabilities.

\section{Problem formulation}
\label{sec:problem}

\subsection{Immersed boundary model}
\label{sec:ib-model}

We consider a closed elastic membrane that encompasses a region of
viscous, incompressible fluid and which is immersed in a domain of
infinite extent containing the same fluid.  At equilibrium, an unforced
membrane takes the form of a pressurised sphere with radius $R$ and
centred at the origin.  Considering the geometry of the equilibrium
state, it is natural to formulate the governing equations in a spherical
coordinate system.  We will therefore introduce coordinates
$(r,\theta,\phi)$, where $r\in[0,\infty)$ is the distance from the
origin, $\theta\in[0,2\upi)$ is the azimuth angle in the horizontal
plane, and $\phi\in[0,\upi]$ is the polar angle measured downward from
the vertical or $z$-axis.  The fluid motion may then be described by the
incompressible Navier-Stokes equations
\begin{align}
  \label{eq:Navier-Stokes}
  \rho \left(\frac{\partial\u}{\partial t} + \u\bcdot\bnabla\u\right)
  &= -\bnabla{p} + \mu\Delta \u + \f, \\
  \label{eq:incompressible}
  \bnabla\bcdot\u &= 0, 
\end{align}
where $\rho$ is the fluid density and $\mu$ is the dynamic viscosity,
both of which are assumed constant.  The membrane moves with the local
fluid velocity according to
\begin{gather}
  \label{eq:dX/dt}
  \frac{\partial \X}{\partial t} = \u(\X,t),
\end{gather}
where $\X(\xi,\eta,t)$ represents the locations of points on the
immersed surface, parameterised by the Lagrangian coordinates
$\xi\in[0,2\upi)$ and $\eta\in[0,\upi]$.  The coordinates $(\xi,\eta)$ are
analogous to the spherical coordinates $(\theta,\phi)$.

An external force 
\begin{gather}
  \f(\x,t) = \iints\F(\X,t)\, \delta(\x-\X) \sin\eta \;\d\xi\; \d\eta,
\end{gather}
arises in the fluid due to the presence of the elastic membrane, and is
written in terms of a force density $\F(\X,t)$ that is integrated
against the Dirac delta function.  As a result, the force is singular
and is supported only on the membrane locations.  The force density may be
expressed as the variational derivative of a membrane elastic energy
functional $E(\X,t)$ \citep{Peskin2002}
\begin{gather*}
  \F(\X,t) = -\frac{\wp E}{\wp \X}, 
\end{gather*}
where we have used $\wp$ to denote the variation of a functional instead
of the more conventional $\delta$, since that is already
reserved for the Dirac delta function.  We choose an energy functional
that incorporates the effect of membrane stretching but ignores any
resistance to shearing or bending motions.  In the interests of
simplicity, we assume the form
\begin{gather*}
  \label{eq:energy-functional}
  E(\X,t) = \half K(t) \iints
  \left(
    \left\|\frac{1}{\sin\eta}\frac{\partial \X}{\partial \xi}\right\|^2 + 
    \left\|\frac{\partial \X}{\partial \eta}\right\|^2  
  \right) \sin\eta \;\d\xi\; \d\eta ,
\end{gather*}
which describes a membrane that would shrink to a point in the absence
of fluid in the interior, but when filled with fluid has an equilibrium
state in which the fluid pressure and membrane forces are in balance.  This
choice of functional was motivated by \citet{Terzopoulos1988} who
simulated deformable sheets in computer graphics applications, and is
also a simplified version of other energy functionals used in
fluid-structure interaction problems \citep{Huang2009}.  Central to this
paper is the specification of the periodically-varying stiffness
\begin{gather}
  K(t) = \sigma(1+2\tau\sin(\omega t)),
\end{gather}
where $\sigma$ is an elastic stiffness parameter, $\omega$ is the
forcing frequency and $\tau$ is the forcing amplitude.  We restrict the
amplitude parameter to $0 \leqslant \tau \leqslant \textstyle\half$,
corresponding to a membrane that resists stretching but not compression.
As a result
\begin{gather}
  \label{eq:force-density}
  \F(\X,t) = K(t) \Delta_\xi \X, 
\end{gather}
where 
\begin{gather}
  \Delta_\xi 
  = \frac{1}{\sin^2\eta} \, \frac{\partial^2}{\partial\xi^2}
  +\frac{1}{\sin\eta}\, \frac{\partial}{\partial\eta}\left(\sin\eta \,
    \frac{\partial}{\partial\eta}\right)
\end{gather}
denotes the angular Laplacian operator in Lagrangian coordinates.

\subsection{Non-dimensionalisation}
\label{sec:nondim}

To simplify the model and the analysis, we first non-dimensionalise the
problem by introducing the following scalings
\begin{align}
  \x = R\, \widetilde{\x}, \quad
  \X = R \, \widetilde{\X}, \quad
  t  = \frac{1}{\omega}\, \widetilde{t}, \quad
  \u = U_c\, \widetilde{\u}, \quad
  p  = P_c \, \widetilde{p},
\end{align}
where the tildes denote non-dimensional quantities and the
characteristic velocity and pressure scales are
\begin{gather*}
  U_c = R \omega \qquad \text{and} \qquad
  P_c = \rho R^2 \omega^2.
\end{gather*}
Substituting the above quantities into the governing equations
\eqref{eq:Navier-Stokes}--\eqref{eq:force-density} yields
\begin{subequations}
  \label{eq:full-equations}
  \begin{align}
    \label{eq:Navier-Stokes-ndim}
    \frac{\partial\widetilde{\u}}{\partial \widetilde{t}} 
    + \widetilde{\u} \bcdot \widetilde{\bnabla} \widetilde{\u}
    &= -\widetilde{\bnabla} \widetilde{p} 
    + \nu \widetilde{\Delta} \widetilde{\u} + \widetilde{\f}, \\
    \label{eq:incompressible-ndim}
    \widetilde{\bnabla} \bcdot \widetilde{\u} &= 0,\\
    \label{eq:dX/dt-ndim}
    \frac{\partial \widetilde{\X}}{\partial \widetilde{t}} 
    &= \widetilde{\u}(\, \widetilde{\X}, \widetilde{t}\,),\\
    \label{eq:force-ndim}
    \widetilde{\f}(\,\widetilde{\x}, \widetilde{t}\,) 
    &= \iints \widetilde{\F}(\, \widetilde{\X}, \widetilde{t}\, )\, 
    \delta(\,\widetilde{\x}-\widetilde{\X}\,) \sin\eta\; \d\xi\; \d\eta, \\
    \label{eq:force-density-ndim}
    \widetilde{\F}(\,\widetilde{\X}, \widetilde{t}\,) 
    &= \widetilde{K}(\,\widetilde{t}\,) \Delta_\xi \widetilde{\X},\\
    \label{eq:stiffness-ndim}
    \widetilde{K}(\,\widetilde{t}\,) &= \kappa(1+2\tau\sin \widetilde{t}\,), 
  \end{align}
\end{subequations}
where we have introduced a dimensionless viscosity (or reciprocal
Reynolds number) 
\begin{gather}
  \nu = \frac{\mu}{\rho R^2\omega} = \frac{1}{Re},
  \label{eq:nu}
\end{gather}
and a dimensionless IB stiffness parameter 
\begin{gather}
  \kappa = \frac{\sigma}{\rho R^3\omega^2}. 
  \label{eq:kappa}
\end{gather}

From this point onward, the tildes are dropped from all 
variables and in the next two sections we perform a number of
simplifications to the governing equations that make them amenable to
Floquet analysis: (a) linearising the equations; (b) expanding the
solution in terms of vector spherical harmonics; and (c) eliminating the
delta function forcing term in lieu of suitable jump conditions across
the membrane.

\subsection{Linearised vector spherical harmonic expansion}
\label{sec:vsh}

Consider a membrane whose shape is a perturbation of the spherical
equilibrium configuration
\begin{gather*}
  \X(\xi,\eta,0) = \big[1 + \epsilon g(\xi,\eta) \big]\, \unitR(\xi,\eta),
\end{gather*}
where $\unitR$ is the radial unit vector, $g(\xi,\eta)$ is some scalar
function, and $|\epsilon| \ll 1$ is a perturbation parameter.  The
equilibrium state is
\begin{gather*}
  \u_0 = \vec{0}, \qquad
  \X_0 = \unitR,  \qquad
  p_0  = 2K(t) \, H(1-r) + p_a,
\end{gather*}
where $H(r)$ is the unit Heaviside step function and $p_a$ represents
some constant ambient pressure.  We then assume a solution in the form
of a regular perturbation expansion
\begin{align*}
  \u &= \u_0 + \epsilon\u_1 + O(\epsilon^2), \\
  p  &= p_0  + \epsilon p_1 + O(\epsilon^2), \\
  \X &= \X_0 + \epsilon\X_1 + O(\epsilon^2).
\end{align*}
Substituting these expressions into the governing equations and
retaining only those terms of $O(\epsilon)$, we obtain the following
system for the first-order quantities:
\begin{subequations}
  \begin{align}
    \frac{\partial\u_1}{\partial t} &= -\bnabla{p_1} + \nu\Delta \u_1 +
    \f_1, \label{eq:lin-stokes}
    \\ 
    \bnabla\bcdot\u_1 &= 0, \label{eq:lin-inc}\\
    \frac{\partial \X_1}{\partial t} &= \u_1(\X_0,t).
    \label{eq:lin-ib}
  \end{align}
\end{subequations}

The stability of the fluid-membrane system may then be determined by
studying solutions of this simpler linear system for the $O(\epsilon)$
variables.  Because of the symmetry in the problem we look for 
solutions in terms of spherical harmonics, which are eigenfunctions of
the angular Laplacian operator $\Delta_\xi$ and form an orthonormal
basis for sufficiently smooth functions of $(\xi,\eta)$.  The normalised
scalar spherical harmonic of degree $m$ and order $k$ is
\begin{gather}
  Y_{m,k}(\theta,\phi) 
  = (-1)^k \sqrt{\frac{2m+1}{4\upi} \frac{(m-k)!}{(m+k)!}\;}
  \; e^{\iim k\theta} \, P_m^k(\cos\phi),
  \label{eq:ssh}
\end{gather}
where $P_m^k$ denotes the associated Legendre polynomial
\citep{Abramowitz1965}.  The natural generalisation to vector-valued
functions (in our case, the fluid velocity and IB position) are the
\emph{vector spherical harmonics or VSH} for which various definitions
have been proposed in the literature
\citep{Morse1953,Hill1954,Barrera1985}.  For example, writing the
velocity field perturbation $\u_1$ in terms of the VSH proposed by
\citet{Hill1954} has the advantage that it fully decouples the
linearised Navier-Stokes equations (i.e., the unsteady Stokes equations
in~\eqref{eq:lin-stokes}).  However, using Hill's basis in the present
context would lead to significant complications later in our analysis.
Therefore, we instead use the VSH basis derived by \citet{Barrera1985}
that is defined in terms of the scalar spherical
harmonics~\eqref{eq:ssh} as follows:
\begin{align*}
  \vshY(\theta,\phi)   &= Y_{m,k}\, \unitR,	
  \\
  \vshPsi(\theta,\phi) &= r\, \bnabla Y_{m,k} 
  = \frac{\iim k}{\sin\phi}\, Y_{m,k}\unitTheta
  + \frac{\partial Y_{m,k}}{\partial\eta} \, \unitPhi,
  \\
  \vshPhi(\theta,\phi) &= \unitR \times \vshPsi
  = \frac{\partial Y_{m,k}}{\partial\eta}\, \unitTheta
  - \frac{\iim k}{\sin\phi}Y_{m,k}\, \skew3\unitPhi,
\end{align*}
where $\unitTheta$ and $\unitPhi$ are the other two unit vectors in
spherical coordinates.  This choice of basis re-introduces a coupling in
the momentum equations but we will see later on that it has the major
advantage of decoupling the jump conditions.  Furthermore, this basis
decomposes vectors into a radial component and two tangential
components, which provides a more intuitive geometric interpretation of
our results.  Finally, we only require the real part of the solution
modes so that, without loss of generality, we can write the velocity,
IB position and pressure variables as
\begin{align*}
  \u_1 &= u^r(r,t)\vshY^c + u^\Psi(r,t)\vshPsi^c + u^\Phi(r,t)\vshPhi^c, \\
  \X_1 &= X^r(t)\vshY^c + X^\Psi(t)\vshPsi^c + X^\Phi(t)\vshPhi^c, \\
  p_1  &= \hat{p}(r,t) Y_{m,k}^c,
\end{align*}
where the superscript $c$ denotes the real (cosine) part of each
spherical harmonic.

\subsection{Jump condition formulation}
\label{sec:jumps}

We next reformulate the equations by eliminating the delta function
forcing term and replacing it with suitable jump conditions across the
membrane, following the approach used by \citet{Lai2001}.  We first
observe that equation \eqref{eq:dX/dt} is a statement that the membrane
must move with the local fluid velocity. Because the membrane is
infinitesimally thin, the velocity must be continuous across the
membrane, which leads immediately to the first jump condition
\begin{gather}
  \label{eq:ujump}
  \jump{\u_1} =\vec{0},
\end{gather}
where the double brackets $\jump{\cdot}$ indicate the jump in a quantity
across the membrane $\Gamma$.  To be more precise, we define
\begin{gather*}
  \jump{\cdot} := \lim_{\gameps\to0}
  \big((\cdot)|_{\x^+} - (\cdot)|_{\x^-}\big),
\end{gather*}
where we have introduced a thin region $\Oe$ surrounding $\Gamma$ that
extends a distance $\gameps$ outwards from either side of the membrane,
with inner surface $\dOe-$ and outer surface $\dOe+$.  The jump in a
quantity at location $\x\in\Gamma$ is then difference between values at
$\x^+ \in \dOe+$ and $\x^- \in \dOe-$, taken in the limit as $\gameps
\to 0$ where both $\x^+,\x^- \to \x$.
\begin{figure}
  \centering
  \includegraphics{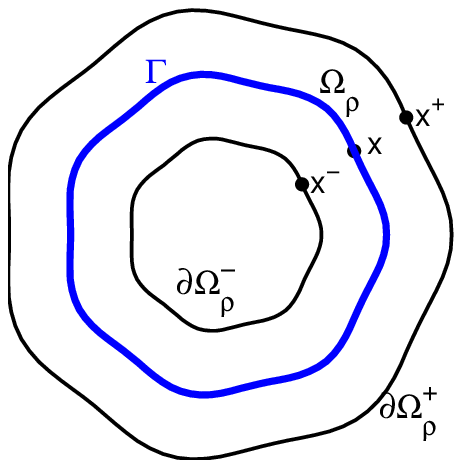}
  \caption{A cross-sectional view of the membrane $\Gamma$ and 
    subdomain $\Oe$, having inner and outer surfaces $\dOe-$ and $\dOe+$
    respectively, which illustrates the limiting process as
    $\x^+,\x^- \to \x$ and $\gameps\to 0$.}
  \label{fig:domains}
\end{figure}

Next, consider the divergence condition \eqref{eq:incompressible-ndim}
which must be satisfied identically on either side of the membrane so
that
\begin{gather}
  \jump{\bnabla\bcdot\u_1} = 0.
\end{gather}
Rewriting this condition in terms of the components of $\u_1$, we
have 
\begin{gather}
  \label{eq:dur-jump}
  \bigjump{\frac{\partial u^r}{\partial r} + \frac{2}{r}\, u^r -
    \frac{m(m+1)}{r}\, u^\Psi} 
  = \bigjump{\frac{\partial u^r}{\partial r}} 
  = 0,
\end{gather}
where the last equality follows from \eqref{eq:ujump}.  

The remaining jump conditions are derived from the momentum equations
\eqref{eq:Navier-Stokes-ndim}.  Letting $\varphi(\x)$ be a smooth test
function with compact support, we multiply the momentum equations by
$\varphi(\x)$ and integrate over $\Oe$ to obtain
\begin{gather}
  \label{eq:subdomain-integral}
  \int_\Oe \left(\frac{\partial\u}{\partial t} + \u\bcdot\bnabla\u
  \right)\varphi(\x) \;\d V   
  = \int_\Oe \left( -\bnabla p + \nu\Delta \u + \f \right)\varphi(\x)\;
  \d V . 
\end{gather}
We now examine each term in this equation in the limit as $\gameps\to
0$. Starting with the left-hand side, we apply the Reynolds transport
theorem to extract the time derivative:
\begin{gather*}
  \int_\Oe \left(\frac{\partial\u}{\partial t} + \u\bcdot\bnabla\u
  \right) \varphi(\x) \; \d V   
  = \frac{\d}{\d t} \int_\Oe \u \varphi(\x)\;  \d V  .
\end{gather*}
Because $\varphi(\x)$ is smooth and $\u$ is bounded, we have that
\begin{gather}
  \lim_{\gameps\to0} \frac{\d}{\d t} \int_\Oe \u \varphi(\x)\;  \d V  = 0.
  \label{eq:limit-dudt}
\end{gather}
Considering next the pressure term in the right-hand side of
\eqref{eq:subdomain-integral}, integrate by parts to obtain
\begin{gather}
  \label{eq:ibp-p}
  \int_\Oe -\bnabla p\, \varphi(\x)\; \d V  
  = -\int_{\dOe+} p \normal\, \varphi(\x) \; \d A  + \int_{\dOe-}
  p\normal\, \varphi(\x) \; \d A
  + \int_\Oe p\, \bnabla\varphi(\x) \; \d V ,
\end{gather}
where $\normal$ is the unit outward normal vector defined by
\begin{gather}
  \label{eq:normal}
  \normal = \frac{\frac{\partial\X}{\partial\eta}\times\frac{\partial\X}{\partial\xi}}
  {\left\| \frac{\partial\X}{\partial\eta}\times\frac{\partial\X}{\partial\xi} \right\|}.
\end{gather}
Because $p$ is also bounded, the last term in \eqref{eq:ibp-p} vanishes as
$\gameps\to0$ and we are left with the difference between two surface
integrals, which reduces to a jump in pressure; in other words,
\begin{gather}
  \int_\Oe -\bnabla p \varphi(\x) \; \d V \quad \to \quad -\int_\Gamma
  \jump{p}\normal \varphi(\x) \; \d A  
  \qquad \mbox{ as } \gameps\to 0.
  \label{eq:limit-p}
\end{gather}
In an analogous manner, the viscous term in
\eqref{eq:subdomain-integral} can be integrated by parts to yield
\begin{gather}
  \int_\Oe \nu\Delta \u \varphi(\x)\;  \d V  \quad \to \quad 
  \int_\Gamma \nu \jump{ \normal\bcdot\bnabla\u } \varphi(\x) \; \d A 
  \qquad \mbox{ as } \gameps\to0.
  \label{eq:limit-viscous}
\end{gather}
Finally, we consider the forcing term in \eqref{eq:subdomain-integral}
and apply the sifting property of the Dirac delta function to obtain
\begin{align}
  \int_\Oe \f \varphi(\x)\;  \d V  
  &= \int_\Oe  \iints  \F \delta(\x-\X) \sin\eta \; \d \xi \;
  \d\eta \; \varphi(\x) \; \d V  , \nonumber \\
  &= \iints  \F  \varphi(\X) \sin\eta \; \d\xi\;  \d\eta .
  \label{eq:limit-f}
\end{align}

The results in \eqref{eq:limit-dudt} and
\eqref{eq:limit-p}--\eqref{eq:limit-f} can then be substituted into
\eqref{eq:subdomain-integral}, after which we take the limit as
$\gameps\to0$ to get
\begin{gather*}
  0=\iints \left( 
    - \jump{p}\normal\left\|
      \frac{\partial\X}{\partial\eta}\times\frac{\partial\X}{\partial\xi}
    \right\|  
    + \nu \jump{ \normal\bcdot\bnabla\u } \left\|
      \frac{\partial\X}{\partial\eta}\times\frac{\partial\X}{\partial\xi}
    \right\| 
    + \F\sin\eta  \right)\varphi(\X) \; \d\xi \; \d\eta,
\end{gather*}
where we have also made use of the identity
\begin{gather*}
  dA = \left\|
    \frac{\partial\X}{\partial\eta}\times\frac{\partial\X}{\partial\xi}
  \right\| \; \d\xi \; \d\eta.
\end{gather*}
Because $\varphi$ is arbitrary and smooth, the integrand must be
identically zero, which yields
\begin{gather}
  \label{eq:jump1}
  0 = - \jump{p}\left(
    \frac{\partial\X}{\partial\eta}\times\frac{\partial\X}{\partial\xi}
  \right) 
  + \nu \bigjump{\left(
      \frac{\partial\X}{\partial\eta}\times\frac{\partial\X}{\partial\xi}
    \right)\bcdot\bnabla\u } 
  + \F\sin\eta ,
\end{gather}
where the normal vector has been replaced using \eqref{eq:normal}.

We now make use of the perturbation expansion for $\X$ to write the
terms arising from the normal vector as
\begin{align*}
  \frac{\partial\X}{\partial\eta}\times\frac{\partial\X}{\partial\xi} 
  &= \frac{\partial}{\partial\eta}\left(\hat{r} + \epsilon\X_1 +
    O(\epsilon^2) \right) 
  \times\frac{\partial}{\partial\xi}\left(\hat{r} + \epsilon\X_1 +
    O(\epsilon^2) \right) ,\\ 
  &= \sin\eta\, \hat{r} + \epsilon\sin\eta\Big[
  (2X^r-m(m+1)X^\Psi)\vshY^c \\ 
  &\qquad\qquad\qquad\qquad +\, (X^\Psi-X^r)\vshPsi^c + X^\Phi\vshPhi^c
  \Big] + O(\epsilon^2). 
\end{align*}
Similarly, the force density can be expanded as
\begin{align*}
  \F &= K(t)\Delta_\xi\left(\hat{r} + \epsilon\X_1+ O(\epsilon^2)
  \right) , \\ 
  &= -2K(t)\hat{r} + \epsilon K(t)\Big[\left(2m(m+1)X^\Psi-(m^2+m+2)X^r
  \right)\vshY^c \\ 
  &\qquad\qquad + \,\left(2X^r - m(m+1)X^\Psi \right)\vshPsi^c -
  m(m+1)X^\Phi\vshPhi^c \Big] + O(\epsilon^2). 
\end{align*}
The remaining jump conditions are obtained by substituting these last
two equations along with the perturbation expansions for $\u$ and $p$
into \eqref{eq:jump1}.  The radial component of \eqref{eq:jump1} gives
two jump conditions for the pressure variables
\begin{align}
  \jump{p_0} &= -2K(t),	\\
  \label{eq:pressure-jump}
  \jump{p_1} &= -K(t) (m-1)(m+2) X^r Y_{m,k}^c, 
\end{align}
while the $\vshPsi$ and $\vshPhi$ components give jump conditions for the
radial derivatives of $u^\Psi$ and $u^\Phi$
\begin{align}
  \label{eq:duPsi-jump}
  \nu\bigjump{\frac{\partial u^\Psi}{\partial r}} &= K(t)(m-1)(m+2) X^\Psi,\\
  \label{eq:duPhi-jump}
  \nu\bigjump{\frac{\partial u^\Phi}{\partial r}} &= K(t)(m-1)(m+2) X^\Phi.
\end{align}
Note that the right-hand sides of each jump condition are completely
decoupled, which is the major advantage to our particular choice of VSH
basis that we referred to earlier in section~\ref{sec:vsh}.

We can now summarise the system of equations that will be analysed in
the remainder of this paper:
\begin{align}
  \label{eq:dur/dt}
  \frac{\partial u^r}{\partial t} &= -\frac{\partial \hat{p}}{\partial r} 
  + \nu\left(\Delta u^r - \frac{2}{r^2}\, u^r + \frac{2m(m+1)}{r^2}\, u^\Psi\right),\\
  \label{eq:duPsi/dt}
  \frac{\partial u^\Psi}{\partial t} &= -\frac{\hat{p}}{r} 
  + \nu\left(\Delta u^\Psi + \frac{2}{r^2}\, u^r\right),\\
  \label{eq:duPhi/dt}
  \frac{\partial u^\Phi}{\partial t} &= \nu\Delta u^\Phi,\\
  \label{eq:divu1}
  \bnabla\cdot\u_1 &= \left( \frac{1}{r^2}\frac{\partial}{\partial r}(r^2u^r) -
  \frac{m(m+1)}{r}\, u^\Psi \right)Y_{m,k}^c = 0,\\  
  \label{eq:dX1/dt}
  \frac{\partial \X_1}{\partial t} &= \u_1|_{r = 1}.
\end{align}
The scalar Laplacian operator simplifies to
\begin{gather*}
  \Delta = \frac{1}{r^2}\frac{\partial}{\partial r}
  \left(r^2\frac{\partial}{\partial r}\right)  
  - \frac{m(m+1)}{r^2}, 
\end{gather*}
and the $O(\epsilon)$ quantities $\X_1$, $\u_1$ and $p_1$ are also
subject to the jump conditions \eqref{eq:ujump}, \eqref{eq:dur-jump} and
\eqref{eq:pressure-jump}--\eqref{eq:duPhi-jump}.  Note that the equation
for $u^\Phi$ is totally decoupled which is another advantage of our
choice of VSH basis.  We also observe that the dynamics of the
linearised solution depend only on the degree $m$ of the spherical
harmonic and not on its order $k$; hence solution modes are
characterised by a single integer $m$.  

In the remainder of this paper, we will drop the subscript ``1'' that
until now has distinguished the $O(\epsilon)$ quantities.

\section{Floquet analysis for an inviscid fluid}
\label{sec:inviscid}

To afford some insight into parametric instabilities occurring in a
simpler version of the immersed membrane problem, we first consider 
an inviscid fluid for which the governing equations reduce to
\begin{subequations}
  \label{eq:inviscid}
  \begin{align}
    \frac{\partial u^r}{\partial t}    &= -\frac{\partial
      \hat{p}}{\partial r}, \label{eq:inviscid-ur}\\
    \frac{\partial u^\Psi}{\partial t} &= -\frac{\hat{p}}{r},\\
    \frac{\partial u^\Phi}{\partial t} &= 0,\\
    \bnabla\bcdot\u &= 0,\\ 
    \frac{\d X^r}{\d t} &= u^r(1,t), \label{eq:inviscid-noflow}\\
    \jump{ u^r }    &= 0,\\
    \jump{\hat{p}}  &= -K(t) (m-1)(m+2) X^r. \label{eq:inviscid-pjump}
  \end{align}
\end{subequations}
Notice that in the absence of viscosity, we are only permitted to impose
the zero normal flow condition \eqref{eq:inviscid-noflow} at the
fluid-membrane interface instead of the usual no-slip condition.

We begin by solving for the pressure away from the membrane, which satisfies
\begin{align}
  \Delta \hat{p} = r^2 \frac{\partial^2 \hat{p}}{\partial r^2} + 2r
  \frac{\partial \hat{p}}{\partial r} - m(m+1) \hat{p} = 0.
  \label{eq:ode-p}
\end{align}
Imposing the requirement that the pressure be bounded at $r=0$ and as
$r\to\infty$ yields 
\begin{align*}
  \hat{p}(r,t) = \begin{cases}
    a(t)\, r^m,      & \mbox{if } r<1, \\
    b(t)\, r^{-m-1}, & \mbox{if } r>1,
  \end{cases}
\end{align*}
where $a(t)$ and $b(t)$ are as-yet undetermined functions of time.
Substituting the pressure solution into the inviscid momentum
equation~\eqref{eq:inviscid-ur} yields the radial fluid acceleration
\begin{align*}
  \frac{\partial u^r}{\partial t} =
  \begin{cases}
      -m a(t)\, r^{m-1},        & \mbox{if } r<1, \\
      (m+1)b(t)\, r^{-m-2}, & \mbox{if } r>1.
  \end{cases}
\end{align*}
Since the fluid cannot pass through the membrane, the radial
acceleration of the fluid and membrane must match and
\begin{gather}
  \frac{\d^2 X^r}{\d t^2} 
  = \left.\frac{\partial u^r}{\partial t}\right|_{r=1}  
  = -m a(t) = (m+1)b(t),
  \label{eq:d2xdt2}
\end{gather}
where the last equality follows from continuity of $u^r$ at the
interface.  This allows us to determine the functions
\begin{gather*}
	a(t) = -\frac{1}{m}\frac{\d^2 X^r}{\d t^2} 
	\qquad \text{and} \qquad
	b(t) = \frac{1}{m+1}\frac{\d^2 X^r}{\d t^2},
\end{gather*}
after which the pressure jump \eqref{eq:inviscid-pjump} may be
expressed as 
\begin{align}
  b(t) - a(t) = -K(t)(m-1)(m+2) X^r(t).
  \label{eq:b-minus-a}
\end{align}
We then have the following equation for the membrane location
\begin{gather}
  \label{eq:Xr-Mathieu}
  \frac{\d^2 X^r}{\d t^2} + \widetilde{\omega}^2
  (1+2\tau\sin t)X^r = 0,
\end{gather}
where 
\begin{gather*}
  \widetilde{\omega}^2 = \frac{\kappa m(m-1)(m+1)(m+2)}{2m+1}.
\end{gather*}
Recalling the definition of $\kappa$ in \eqref{eq:kappa}, we may write
$\widetilde{\omega}=\omega_N/\omega$ as the ratio of the natural
frequency $\omega_N$ for the unforced problem to the forcing frequency
$\omega$, where
\begin{gather}
  \label{eq:omegaN}
  \omega_N^2 = \frac{\sigma m(m-1)(m+1)(m+2)}{\rho R^3 (2m+1)}.
\end{gather}
This natural oscillation frequency matches exactly with the classical
result of \citet[Art.~253]{Lamb1932} for oscillations of a spherical
liquid drop, with the only difference being that a surface tension force
replaces the elastic restoring force in our IB context.

We now focus on \eqref{eq:Xr-Mathieu} which describes a harmonic
oscillator with a time-dependent frequency parameter, and hence takes
the form of the well-known Mathieu equation~\citep{Nayfeh1993}.  We
invoke the Floquet theory and look for a series solution of the form
\begin{gather*}
  X^r(t) = e^{\gamma t} \sum_{n = -\infty}^{\infty} X^r_n e^{\iim n t},
\end{gather*}
where $\gamma \in \mathbb{C}$ is the Floquet exponent that determines
the stability of solution as $t \to \infty$.  Substituting this series
expansion into \eqref{eq:Xr-Mathieu} yields the following infinite
system of linear algebraic equations:
\begin{gather}
  \label{eq:Floquet-inviscid}
  -\iim\left[\frac{(\gamma + \iim n)^2}{\widetilde{\omega}^2} +
    1\right]X^r_n 
  = \tau (X^r_{n-1} - X^r_{n+1})
  \qquad
  \text{for } n \in \mathbb{Z}.
\end{gather}

Our aim is now to investigate all non-trivial solutions of this system
and determine what (if any) instabilities can arise, and for what
parameter values they occur.  To this end, we use the standard approach
and restrict ourselves to periodic solutions having $\Real\{\gamma\}=0$
(i.e., neutral stability) and corresponding to two special values of the
Floquet exponent: $\gamma = 0$ (harmonic modes) and $\textstyle{\gamma =
  \half \iim}$ (subharmonic modes).  It is possible to show that
all other values of $\Imag\{\gamma\}$ lead to modes that decay in time.
To ensure that the resulting solutions are real-valued functions, we
impose a set of \emph{reality conditions} on the Fourier series
coefficients, which for the harmonic case are
\begin{gather}
  \label{eq:reality-harmonic}
  X^r_{-n} = \overline{X}^r_{n},
\end{gather}
whereas for the subharmonic case
\begin{gather}
  \label{eq:reality-subharmonic}
  X^r_{-n} = \overline{X}^r_{n-1},
\end{gather}
for all $n\in\mathbb{Z}$.  Either set of reality conditions implies that
it is only necessary to consider non-negative values of $n$, and so the
linear system \eqref{eq:Floquet-inviscid} may be written as
\begin{gather}
  \label{eq:inviscid-system}
  A_n X^r_n = \tau (X^r_{n-1} - X^r_{n+1}), 
\end{gather}
now with $n = 0,1,2,\dots$.  For the purposes of our stability analysis,
the forcing amplitude $\tau$ is treated as an unknown, and in practice,
we approximate the infinite linear system by truncating at a finite
number of modes $N$.  The resulting equations may be written compactly
in matrix form as
\begin{gather}
  \label{eq:gen-eig}
  \mat{A} \vec{v} = \tau \mat{B} \vec{v}, 
\end{gather}
where the unknown series coefficients are collected together in a vector
\begin{gather*}
  \vec{v} = 
  \left[
    \begin{array}{c}
      \vdots	\\
      \Real\{X^r_n\}	\\
      \Imag\{X^r_n\}	\\
      \vdots
    \end{array}
  \right],
\end{gather*}
and $\mat{A} = \mbox{diag}(\mat{A}_0,\mat{A}_1,\ldots,\mat{A}_N)$ is a
block diagonal matrix with $2\times 2$ blocks
\begin{gather*}
  \mat{A}_n =
  \left[
    \begin{array}{rr}
      \Real\{A_n\} &  -\Imag\{A_n\} \\
      \Imag\{A_n\} & \Real\{A_n\}  \\
    \end{array}
  \right].
\end{gather*}
Similarly, $\mat{B}$ has a block tridiagonal structure of the form
\begin{gather*}
  \mat{B} = 
  \left[
    \begin{array}{rrrrr}
      \skew3\widehat{\mat{B}} & \skew3\widetilde{\mat{B}} & & &\\
      \mat{I}_2   & \mat{0}_2 & -\mat{I}_2 & & \\
      & \mat{I}_2 & \mat{0}_2 & -\mat{I}_2 &\\
      & & \ddots  &  \ddots   & \ddots 
    \end{array}
  \right],
\end{gather*}
where $\mat{0}_2$ and $\mat{I}_2$ denote the $2\times 2$ zero and identity
matrices respectively.  The sub-matrices making up the first block rows
of $A$ and $B$ depend on the value of $\gamma$, so that for the harmonic
case (with $\gamma=0$)
\begin{gather*}
  \mat{A}_0 = I, 
  \qquad
  \skew3\widehat{\mat{B}} = \left[
    \begin{array}{rr}
      0 & 2	\\
      0 & 0
    \end{array}
  \right],
  \qquad	
  \skew3\widetilde{\mat{B}}= \mat{0}_2,
\end{gather*}
while for the subharmonic case (with $\gamma=\textstyle\half\iim$)
\begin{gather*}
  \mat{A}_0 = 
  \left[
    \begin{array}{rr}
      \Real\{A_0\} &  -\Imag\{A_0\} \\
      \Imag\{A_0\} & \Real\{A_0\}  \\
    \end{array}
  \right], 
  \qquad
  \skew3\widehat{\mat{B}} = \left[
    \begin{array}{rr}
      1 & 0	\\
      0 & -1
    \end{array}
  \right] ,
  \qquad	
  \skew3\widetilde{\mat{B}} = -\mat{I}_2.
\end{gather*}
Equation \eqref{eq:inviscid-system} can be viewed as a generalised
eigenvalue problem with eigenvalue $\tau$ and eigenvector $\vec{v}$.
Therefore, determining the stability of the parametrically forced
spherical membrane reduces to finding all values of $\tau$ and $\vec{v}$
for the two Floquet exponents $\gamma=0$ and $\textstyle{\half
  \iim}$ (corresponding to the harmonic and subharmonic cases
respectively).

\begin{figure}
  \centering
  \begin{multicols}{2}
  	\includegraphics[height=0.85\textheight]{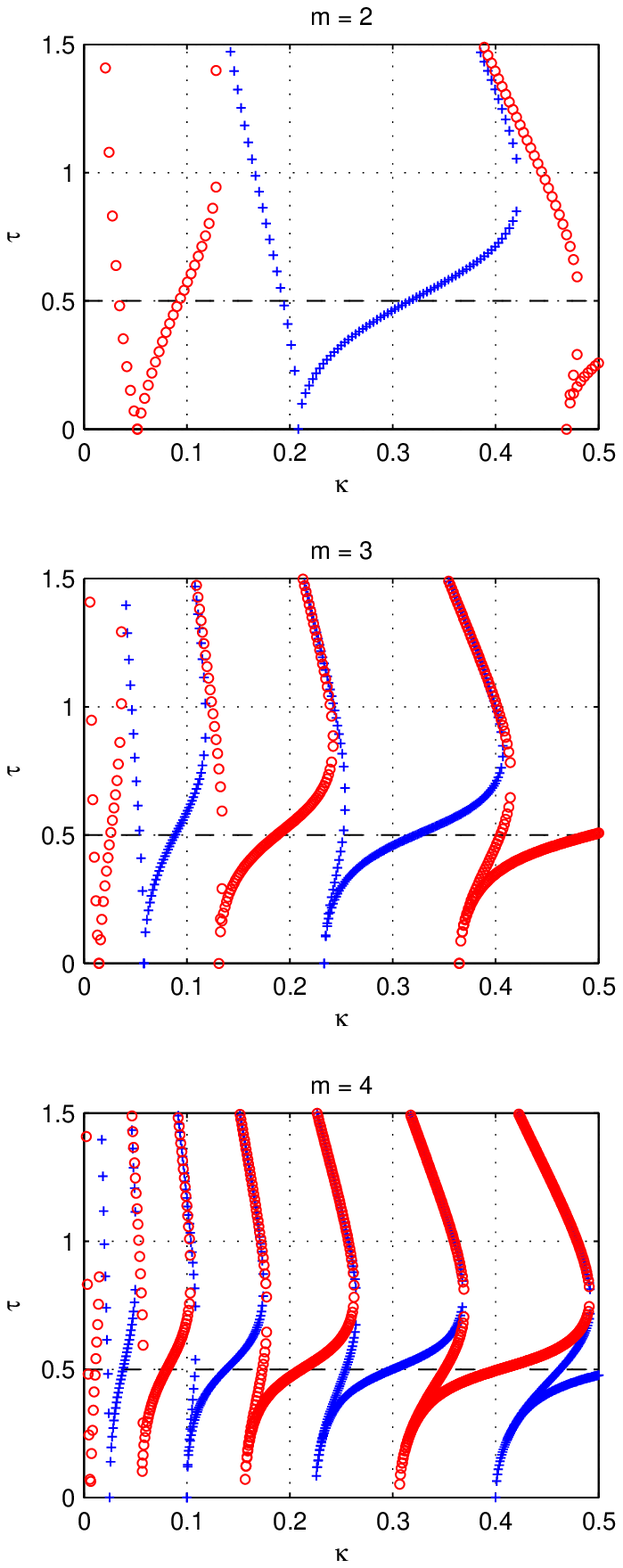}
  \columnbreak
  	\\
	\hspace{0.4cm}
  	\includegraphics[height=0.837\textheight]{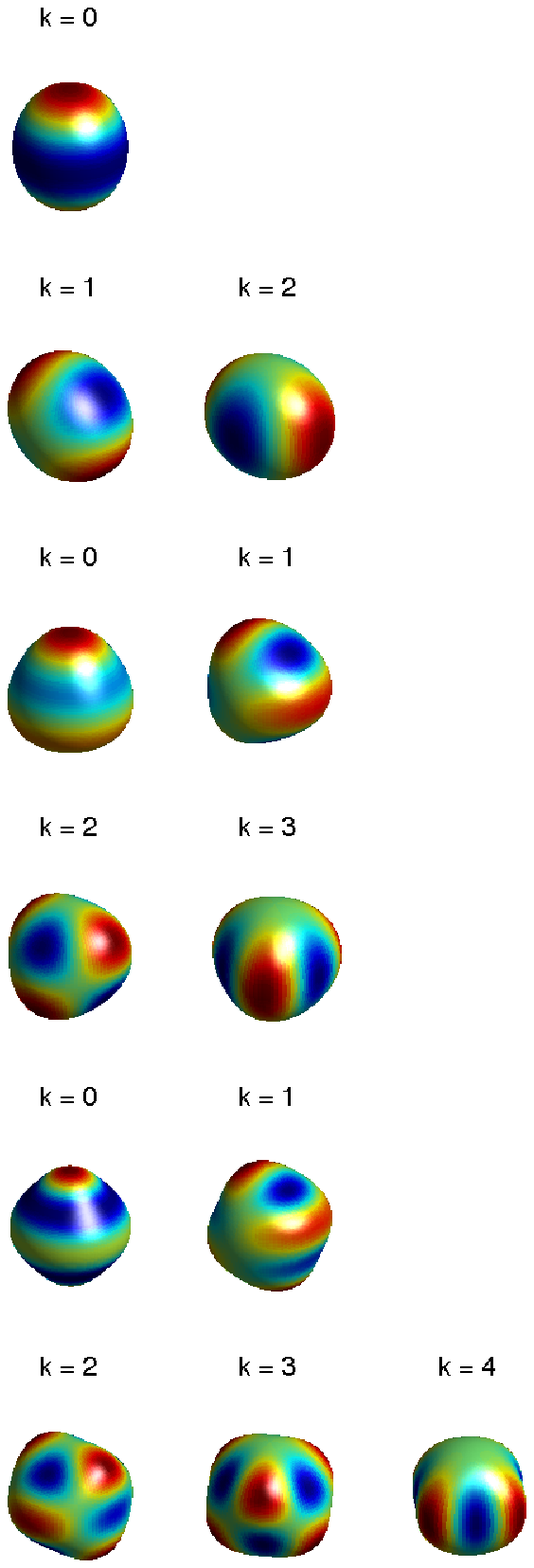}
  \end{multicols}
  \caption{On the left are Ince-Strutt diagrams for the inviscid
    problem, showing stability contours for spherical harmonics $m=2$
    (top), 3 (middle) and 4 (bottom).  Subharmonic modes are denoted
    with {\color{Red}$\bigcirc$} and harmonic modes with {\color{Blue}$+$},
    and only modes with $\textstyle\tau<\half$ are physically
    relevant.  To the right of each Ince-Strutt diagram is a picture
    showing spherical shells perturbed non-negative $k$-modes.}
  \label{fig:contours-inviscid}
\end{figure}

A particularly effective way of visualising these solutions is to vary
one of the system parameters (either forcing frequency or elastic
stiffness) and to consider the \emph{stability regions} that are
generated as the eigenvalues trace out curves in parameter space.  This
diagram is referred to as an \emph{Ince-Strutt
  diagram}~\citep{Nayfeh1993}, three of which are depicted in
figure~\ref{fig:contours-inviscid} as plots of $\kappa$ versus $\tau$
for three different spherical harmonics, $m=2$, 3 and 4.  The stability
boundaries take the form of \emph{fingers} or \emph{tongues} that extend
downward in parameter space.  There are clearly two distinct sets of
alternating fingers corresponding to harmonic and subharmonic modes,
which we denote using the two point types {\color{Blue}$+$} and
{\color{Red}$\bigcirc$} respectively.  Parameter values lying above and
inside any given finger correspond to unstable solutions, whereas all
parameters lying below the fingers correspond to stable solutions.  It
is essential to keep in mind that only parameters lying below the
horizontal line ${\textstyle \tau=\half}$ are physically relevant,
since these values of $\tau$ correspond to a membrane stiffness $K(t)$
that remains positive.

It is evident from these diagrams that for a given forcing amplitude
$\tau$, an immersed spherical shell can experience parametric
instability for a disjoint set of $\kappa$ ranges (with corresponding
ranges of the physical parameters $\omega$, $\rho$, $R$ and $\sigma$
according to \eqref{eq:kappa}).  For example, different unstable modes
(corresponding to different integer values of $m$) can be excited by
forcing the system within a given range of $\omega$.  Furthermore, there
are an infinite number of unstable modes that can be excited since the
harmonic and subharmonic fingers continue to alternate to the right
forever as $\kappa$ increases.

These membrane instabilities exist for all values of forcing amplitude
$0< \textstyle\tau<\half$ since each of the stability
fingers extends downward and touches the $\kappa$-axis as
$\tau\to 0$.  The points at which the fingers touch down correspond to
the natural oscillation frequencies $\omega_N$ for the unforced problem
given in \eqref{eq:omegaN}.  Indeed, we see that for $|\tau| \ll 1$,
resonances occur at forcing frequencies
\begin{gather*}
  \frac{\omega_N}{\omega} = \frac{\ell}{2},
\end{gather*}
for any positive integer $\ell$; that is, the natural frequency
for a given $m$-mode is an integer multiple of half the parametric
forcing frequency. Parametric resonance is characterised by
the subharmonic modes (odd $\ell$)
where the response frequency is half the forcing frequency.

\section{Floquet analysis for a viscous fluid} 
\label{sec:viscous}

Guided by the analysis from the previous section, we next apply Floquet
theory to the original governing equations with viscosity by looking for
series solutions of the form 
\begin{align*}
  \u(r,\theta,\phi,t) &= e^{\gamma t} \sum_{n=-\infty}^{\infty} e^{\iim nt}\left(
    u^r_n(r)\vshY^c + u^\Psi_n(r)\vshPsi^c + u^\Phi_n(r)\vshPhi^c\right),\\
  \X(\xi,\eta,t)      &= e^{\gamma t} \sum_{n=-\infty}^{\infty} e^{\iim nt}\left(
    X^r_n\vshY^c + X^\Psi_n\vshPsi^c + X^\Phi_n\vshPhi^c\right),\\
  p(r,\theta,\phi,t)  &= e^{\gamma t} \sum_{n=-\infty}^{\infty}e^{\iim nt} p_n(r) Y_{m,k}^c.
\end{align*}
The pressure coefficients $p_n(r)$ satisfy the same ODE~\eqref{eq:ode-p}
and boundedness conditions as in the inviscid case, and so the solution
has the same form
\begin{align*}
  p_n(r) =
  \begin{cases}
    a_{n} r^m,      & \mbox{if } r<1, \\[2pt]
    b_{n} r^{-m-1}, & \mbox{if } r>1, \\
  \end{cases}
\end{align*}
with the only difference being that $a_n$ and $b_n$ are constants.
Next, combining the radial momentum equation \eqref{eq:dur/dt} with the
divergence-free condition \eqref{eq:divu1} yields the PDE
\begin{align*}
  \frac{\partial u^r}{\partial t}
  = -\frac{\partial \hat{p}}{\partial r} 
  + \nu\left(
    \Delta u^r - 2\frac{u^r}{r^2} 
    + \frac{2}{r}\left(\frac{1}{r^2}\frac{\partial}{\partial r}(r^2 u^r)\right)
  \right).
\end{align*}
After substituting the series form for $u^r$ we obtain 
\begin{align*}  
  r^2 {u^r_n}'' + 4r{u^r_n}' + \left(2 - m(m+1) -
    \beta_n^2 r^2 \right)u^r_n - \frac{r^2}{\nu}\, p_n' = 0, 
\end{align*}
where primes denote derivatives with respect to $r$ and
\begin{gather*}
  \beta_n = \sqrt{\frac{\gamma + \iim n}{\nu}}
  \quad 
  \mbox{ with }
  \quad
  \Real\{\beta_n\}>0, 
\end{gather*}

We first consider the situation where the quantity $\beta_n \ne 0$, in
which case the radial velocity component can be expressed as
\begin{gather}
  u^r_n(r) = \int_0^\infty \frac{z^2}{\nu} \, G(r,z) p_n'(z) \; \d z,
  \label{eq:ur-integral}
\end{gather}
where $G(r,z)$ is the Green's function satisfying
\begin{gather*}
  r^2\frac{\partial^2 G}{\partial r^2} + 4r\frac{\partial G}{\partial r}
  + (2-m(m+1) - \beta_n^2r^2)G = \delta(r-z),
\end{gather*}
along with the two jump conditions
\begin{gather*}
  \left.G\right|_{r=z^-}^{r=z^+} = 0 
  \qquad \text{and} \qquad
  \left.r^2\frac{\partial G}{\partial r}\right|_{r=z^-}^{r=z^+} = 1.
\end{gather*}
The Green's function can be obtained explicitly as
\begin{gather*}
  G(r,z) =
  \begin{cases}
    \displaystyle
    z\beta_n h_m(\iim\beta_nz)\, \frac{j_m(\iim\beta_nr)}{r},
    &\quad\mbox{ if } r < z,\\[0.3cm] 
    \displaystyle
    z\beta_n j_m(\iim\beta_nz)\, \frac{h_m(\iim\beta_nr)}{r},
    &\quad\mbox{ if } r > z, 
  \end{cases}
\end{gather*}
where $j_m$ and $h_m$ denote the $m$th-order spherical Bessel and Hankel
functions of the first kind (respectively).  The expression in
\eqref{eq:ur-integral} can be integrated explicitly to obtain the radial
velocity as
\begin{gather*}
  u^r_n(r) = \left\{ 
    \begin{array}{lll}
      &\displaystyle
      -\frac{\iim j_m(\iim\beta_nr)}{\nu r}
      \Big(m a_n h_{m+1}(\iim\beta_n) - (m+1) b_n
      h_{m-1}(\iim\beta_n)\Big)\\[0.2cm]
      &\displaystyle\qquad\qquad
      -\frac{m}{\nu\beta_n^2}a_nr^{m-1},
      &\mbox{ if } r < 1, \\[0.3cm]
      &\displaystyle
      -\frac{\iim h_m(\iim\beta_nr)}{\nu r}
      \Big(m a_n j_{m+1}(\iim\beta_n) - (m+1) b_n
      j_{m-1}(\iim\beta_n)\Big)\\[0.2cm]
      &\qquad\qquad+\frac{m+1}{\nu\beta_n^2} b_n r^{-m-2},
      &\mbox{ if } r > 1.
    \end{array}\right.
\end{gather*}
It is then straightforward to show that radial velocity coefficients are
continuous across the membrane and satisfy
\begin{gather*}
  u_{n}^r(1^+) = u_{n}^r(1^-) 
  = -\frac{\iim m}{\nu}\,   a_{n}h_m(\iim\beta_n)j_{m+1}(\iim\beta_n)
  + \frac{\iim(m+1)}{\nu}\, b_{n}j_m(\iim\beta_n)h_{m-1}(\iim\beta_n).
\end{gather*}
The coefficient $u^\Psi_n$ can be obtained by solving the
incompressibility condition 
\begin{gather*}
  \left(\frac{1}{r^2}\frac{\d}{\d r}\left(r^2 u^r_n\right) 
    - \frac{m(m+1)}{r}u^\Psi_n \right) Y_{m,k}^c = 0, 
\end{gather*}
so that
\begin{align*}
  u^\Psi_n(r) 
  &= \frac{1}{rm(m+1)}\frac{\d}{\d r}\left(r^2 u^r_n\right),\\
  &= \left\{
    \begin{array}{ll}
      &-\frac{\iim}{\nu}\left(\frac{j_m(\iim\beta_nr)}{r} + \iim \beta_n j_m'(\iim\beta_n r)\right)
      \left(a_n \frac{h_{m+1}(\iim\beta_n)}{m+1} - b_n \frac{h_{m-1}(\iim\beta_n)}{m}\right)\\
      &\qquad\qquad -\frac{1}{\nu\beta_n^2} \, a_nr^{m-1},
      \qquad\qquad\qquad\qquad\qquad\qquad\qquad\quad\;\;\;\mbox{ if } r < 1,\\[0.3cm]
      &-\frac{\iim}{\nu}\left(\frac{h_m(\iim\beta_nr)}{r} + \iim \beta_n h_m'(\iim\beta_n r)\right)
      \left(a_n \frac{j_{m+1}(\iim\beta_n)}{m+1} - b_n \frac{j_{m-1}(\iim\beta_n)}{m}\right)\\
      &\qquad\qquad -\frac{1}{\nu\beta_n^2}\, b_nr^{-m-2},
      \qquad\qquad\qquad\qquad\qquad\qquad\qquad\quad\;\;\mbox{ if } r > 1.
    \end{array}
  \right.
\end{align*}
The analogous equation for the third velocity coefficient $u^\Phi_n(r)$
is 
\begin{gather*}
  r^2{u^\Phi_n}'' + 2r {u^\Phi_n}' -  (\beta_n^2r^2 +m(m+1)){u^\Phi_n} = 0,
\end{gather*}
which can be solved to obtain
\begin{gather*}
  u^\Phi_n(r) = 
  \begin{cases}
    c_n j_m(\iim\beta_nr), &\mbox{if } r < 1,\\[0.2cm]
    c_n \, \frac{j_m(\iim\beta_n)}{h_m(\iim\beta_n)}\,
    h_m(\iim\beta_nr), &\mbox{if } r > 1,
  \end{cases}
\end{gather*}
where $c_n$ are arbitrary constants.

The next major step is to determine values of the constants $a_n$, $b_n$
and $c_n$ by imposing the interface conditions \eqref{eq:dX1/dt}.  By
orthogonality, the radial coefficients for the membrane position satisfy
\begin{align*}
  (\gamma + \iim n)X^r_n &= u^r_n(1),\\
  &= -\frac{\iim m}{\nu} \, a_n h_m(\iim\beta_n) j_{m+1}(\iim\beta_n)
  + \frac{\iim(m+1)}{\nu} \, b_n j_m(\iim\beta_n) h_{m-1}(\iim\beta_n),\\
  \intertext{with similar expressions holding for the other two sets of
    coefficients} 
  (\gamma + \iim n)X^\Psi_n  
  &= -\frac{\iim}{(m+1)\nu}\, a_{n} \Big( h_m(\iim\beta_n) 
  + \iim\beta_n h_m'(\iim\beta_n) \Big) j_{m+1}(\iim\beta_n)\\ 
  &  \qquad + \frac{\iim}{m\nu} \, b_{n} \Big( j_m(\iim\beta_n) +
  \iim\beta_nh_m'(\iim\beta_n) \Big) h_{m-1}(\iim\beta_n),\\
  (\gamma + \iim n)X^\Phi_n &= c_n j_m(\iim\beta_n).
\end{align*}
These three equations may then be solved to obtain
\begin{subequations}\label{eq:an-to-cn}
  \begin{align}
    a_n &= 
    - \frac{\iim\nu^2\beta_n^3}{m} \, \frac{j_m(\iim\beta_n) +
      \iim\beta_nj_m'(\iim\beta_n)}{j_{m+1}(\iim\beta_n)}\,X^r_n 
    + \iim\nu^2\beta_n^3(m+1) \,
    \frac{j_m(\iim\beta_n)}{j_{m+1}(\iim\beta_n)}\,X^\Psi_n,
    \label{eq:an}\\  
    b_n &= 
    -\frac{\iim\nu^2\beta_n^3}{m+1} \, \frac{h_m(\iim\beta_n) +
      \iim\beta_n \, h_m'(\iim\beta_n)}{h_{m-1}(\iim\beta_n)}\,X^r_n 
    +\iim\nu^2\beta_n^3 m\,
    \frac{h_m(\iim\beta_n)}{h_{m-1}(\iim\beta_n)}\,X^\Psi_n,
    \label{eq:bn}\\
    c_n &=
    \frac{\nu\beta_n^2}{j_m(\iim\beta_n)} \, X^\Phi_n.
    \label{eq:cn}
\end{align}
\end{subequations}

We are now prepared to impose the jump conditions \eqref{eq:duPsi-jump}
and \eqref{eq:duPhi-jump}, which yield
\begin{align*}
  \nu\left(-\frac{1}{m+1}a_n - \frac{1}{m}b_n \right)
  &= \kappa(m-1)(m+2)\left(X^\Psi_n - \iim\tau X^\Psi_{n-1} 
    + \iim\tau X^\Psi_{n+1}\right), \\
  \nu\left(\iim\beta_n \frac{j_m(\iim\beta_n)}{h_m(\iim\beta_n)} 
    h_m'(\iim\beta_n) -\iim\beta_nj_m'(\iim\beta_n)\right)c_n
  &= \kappa (m-1)(m+2) \left(X^\Phi_n - \iim\tau X^\Phi_{n-1} 
    + \iim\tau X^\Phi_{n+1}\right). 
\end{align*}
After replacing $a_n$, $b_n$ and $c_n$ in these last two expressions
with \eqref{eq:an-to-cn}, we then obtain the following two linear
systems of equations relating the coefficients $X^r_n$, $X^\Psi_n$ and
$X^\Phi_n$:
\begin{multline}
  \label{eq:viscous-system-1}
  \frac{\beta_n^3}{(m-1)(m+2)}\frac{\nu^2}{\kappa}\left(
    \frac{h_m(\iim\beta_n)}{(m+1)h_{m-1}(\iim\beta_n)} -
    \frac{j_m(\iim\beta_n)}{m j_{m+1}(\iim\beta_n)} 
  \right) X^r_n\\
  + \left[ \frac{\beta_n^3}{(m-1)(m+2)}\frac{\nu^2}{\kappa}\left(
      \frac{h_m(\iim\beta_n)}{h_{m-1}(\iim\beta_n)}
      + \frac{j_m(\iim\beta_n)}{j_{m+1}(\iim\beta_n)}
    \right)
    - \iim
  \right]X^\Psi_n\\
  = \tau \left(X^\Psi_{n-1}-X^\Psi_{n+1}\right),
\end{multline}
\begin{gather}
  \label{eq:viscous-system-2}
  \left[\frac{\iim\beta_n}{(m-1)(m+2)
      j_m(\iim\beta_n)h_m(\iim\beta_n)}\frac{\nu^2}{\kappa} -
    \iim\right]X^\Phi_n 
  = \tau\left(X^\Phi_{n-1} - X^\Phi_{n+1}\right).
\end{gather}
In a similar manner, the final jump condition for the pressure
\eqref{eq:pressure-jump} yields
\begin{multline}
  \label{eq:viscous-system-3}
  \left[-\frac{\iim\beta_n^4}{(2m+1)(m-1)(m+2)}\frac{\nu^2}{\kappa}\left(
      2 - \frac{m}{m+1}\frac{h_{m+1}(\iim\beta_n)}{h_{m-1}(\iim\beta_n)}
      - \frac{m+1}{m}\frac{j_{m-1}(\iim\beta_n)}{j_{m+1}(\iim\beta_n)}
    \right)
    -\iim
  \right]X^r_n \\
  -\frac{\iim\beta_n^4}{(2m+1)(m-1)(m+2)}\frac{\nu^2}{\kappa}\left(
    1 - m\frac{h_{m+1}(\iim\beta_n)}{h_{m-1}(\iim\beta_n)}
    + (m+1)\frac{j_{m-1}(\iim\beta_n)}{j_{m+1}(\iim\beta_n)}
  \right)X^\Psi_n\\
  = \tau\left(X^r_{n-1}-X^r_{n+1}\right).
\end{multline}
When taken together, equations
\eqref{eq:viscous-system-1}--\eqref{eq:viscous-system-3} represent a
solvable system for $X^r_n$, $X^\Psi_n$ and $X^\Phi_n$ in the case when
$\beta_n\neq 0$.


We now consider the special case $\beta_n = 0$, which occurs only when
$n=0$ and $\gamma=0$ (i.e., for harmonic modes) and the equation for
radial velocity reduces to
\begin{gather*}
  r^2 {u^r_0}'' + 4r{u^r_0}' + \left(2 - m(m+1)\right)u^r_0 -
  r^2\frac{p_0'}{\nu} = 0. 
\end{gather*}
The corresponding Green's function is
\begin{gather*}
  G(r,z) = -\frac{1}{2m+1} 
  \begin{cases}
    \frac{r^{m-1}}{z^{m}},   &\mbox{ if } r < z,\\[8pt]
    \frac{z^{m+1}}{r^{m+2}}, &\mbox{ if } r > z,
  \end{cases}
\end{gather*}
from which we obtain 
\begin{gather}
  u^r_0(r) = \begin{cases}
      a_0\frac{m }{\nu(2m+3)} r^{m+1} -
      \frac{1}{\nu(2m+1)}\left(a_0\frac{m}{2} - b_0\frac{m+1}{2m-1}
      \right)r^{m-1}, 
      &\mbox{ if } r<1, \\[8pt]
      b_0\frac{m+1}{\nu(2m-1)} r^{-m} -
      \frac{1}{\nu(2m+1)}\left(a_0\frac{m}{2m+3} +
        b_0\frac{m+1}{2}\right)r^{-m-2}, 
      &\mbox{ if } r>1.
  \end{cases}
  \label{eq:u0-r}
\end{gather}
Using the incompressibility condition as before, we find that 
\begin{gather}
  u^\Psi_0(r) = \begin{cases}
    a_0\frac{m+3}{\nu(2m+3)(m+1)} \, r^{m+1} \\
    \qquad\qquad - \, \frac{1}{\nu m(2m+1)} \, \left(a_0\frac{m}{2} - 
      b_0\frac{m+1}{2m-1} \right)r^{m-1}, 
    & \mbox{ if } r<1, \\[8pt]
    -b_0\frac{m-1}{\nu m(2m-1)} \, r^{-m} \\
    \qquad\qquad + \, \frac{1}{\nu(m+1)(2m+1)} \,
    \left(a_0\frac{m}{2m+3} +  b_0\frac{m+1}{2}\right)r^{-m-2}, 
    & \mbox{ if } r>1,
  \end{cases}
  \label{eq:u0-Psi}
\end{gather}
and remaining velocity coefficients are given very simply by
\begin{align}
  u^\Phi_0(r) = \begin{cases}
    c_0 r^m,	  & \mbox{if } r<1, \\[2pt]
    c_0 r^{-m-1}, & \mbox{if } r>1. \\
  \end{cases}
  \label{eq:u0-Phi}
\end{align}
If we then substitute \eqref{eq:u0-r}--\eqref{eq:u0-Phi} into the
membrane evolution equation \eqref{eq:dX1/dt}, we obtain the following
system of three linear equations
\begin{align*}
  -\frac{m}{\nu(2m+3)(2m+1)}\, a_0 + \frac{m+1}{\nu (2m-1)(2m+1)}\,
  b_0 &= 0,\\
  \frac{m}{\nu (m+1)(2m+3)(2m+1)}\, a_0 + \frac{m+1}{\nu
    m(2m-1)(2m+1)}\, b_0 &= 0, \\
  c_0 &= 0.
\end{align*}
It is easy to show that this linear system is invertible as long as $m$
is a positive integer, and since the equations are homogeneous the
unique solution is $a_0 = b_0 = c_0 = 0$.  Therefore, in the special
case $\beta_n=0$, the jump conditions
\eqref{eq:viscous-system-1}--\eqref{eq:viscous-system-3} reduce to
\begin{align}
  \label{eq:beta0-1}
  X^\Psi_n -\iim\tau X^\Psi_{n-1}+ \iim\tau X^\Psi_{n+1} &= 0,\\
  \label{eq:beta0-2}
  X^r_n - \iim\tau X^\Phi_{n-1}+ \iim\tau X^\Phi_{n+1} &= 0, \\
  \label{eq:beta0-3}
  X^r_n  - \iim\tau X^r_{n-1} + \iim\tau X^r_{n+1} &= 0.
\end{align}


To investigate the stability of a parametrically-forced elastic shell,
we now need only consider the $\vshY$ and $\vshPsi$ solution components.
This is because the $\vshPhi$ component is completely decoupled in the
momentum equations and neither is it driven by a pressure gradient, so
that it evolves independently (this is of course closely related to the
fact that the equations \eqref{eq:viscous-system-2} for $X^\Phi_n$ are
decoupled from the other equations).  In particular, if the $\vshPhi$
component of the initial membrane position is zero, then it will remain
zero for all time.  As a result, it is only necessary to consider
equations \eqref{eq:viscous-system-1}, \eqref{eq:viscous-system-3},
\eqref{eq:beta0-1} and \eqref{eq:beta0-3}, which can be written as
\begin{align}
  A_n X^r_n + B_n X^\Psi_n &= \tau (X^r_{n-1} - X^r_{n+1}),
  \label{eq:viscous-final-1}\\
  C_n X^r_n + D_n X^\Psi_n &= \tau (X^\Psi_{n-1} - X^\Psi_{n+1}),
  \label{eq:viscous-final-2}
\end{align}
for suitably defined constants $A_n$, $B_n$, $C_n$ and $D_n$.  We again
impose reality conditions for either harmonic solutions
($\gamma=0$)
\begin{subequations}
  \label{eq:reality-harmonic-2}
  \begin{align}
    X^r_{-n}    &= \overline{X}^r_{n},\\
    X^\Psi_{-n} &= \overline{X}^\Psi_{n},
  \end{align}
\end{subequations}
or subharmonic solutions ($\gamma=\textstyle\half\iim$)
\begin{subequations}
  \label{eq:reality-subharmonic-2}
  \begin{align}
    X^r_{-n}    &= \overline{X}^r_{n-1}, \\
    X^\Psi_{-n} &= \overline{X}^\Psi_{n-1}.
  \end{align}
\end{subequations}
Again, we only need to consider non-negative integer values of
$n=0,1,\dots,N$, so that equations
\eqref{eq:viscous-final-1}--\eqref{eq:viscous-final-2} take the form of
a generalised eigenvalue problem $\mat{A} \vec{v} = \tau \mat{B}
\vec{v}$, where the solution vector
\begin{gather*}
  \vec{v} = 
  \left[
    \begin{array}{c}
      \vdots	\\
      \Real\{X^r_n\}	\\[2pt]
      \Imag\{X^r_n\}	\\[2pt]
      \Real\{X^\Psi_n\}	\\[2pt]
      \Imag\{X^\Psi_n\}	\\
      \vdots
    \end{array}
  \right].
\end{gather*}
is of length $4(N+1)$.  The matrix $\mat{A} = \mbox{diag}(\mat{A_0},
\mat{A_1}, \ldots, \mat{A_N})$ is block diagonal consisting of $4\times
4$ blocks
\begin{gather*}
  \mat{A}_n =
  \left[
    \begin{array}{rrrr}
      \Real\{A_n\} & -\Imag\{A_n\} & \Real\{B_n\} & -\Imag\{A_n\}	\\[2pt]
      \Imag\{A_n\} &  \Real\{A_n\} & \Imag\{B_n\} &  \Real\{B_n\} 	\\[2pt]
      \Real\{C_n\} & -\Imag\{C_n\} & \Real\{D_n\} & -\Imag\{D_n\}	\\[2pt]
      \Imag\{C_n\} &  \Real\{C_n\} & \Imag\{D_n\} &  \Real\{D_n\} 	\\
    \end{array}
  \right],
\end{gather*}
and $\mat{B}$ is a block tridiagonal matrix of the form
\begin{gather*}
  \mat{B} = 
  \left[
    \begin{array}{rrrrr}
      \skew3\widehat{\mat{B}} & \skew3\widetilde{\mat{B}} & & &\\
      \mat{I}_4    & \mat{0}_4 & -\mat{I}_4 & & \\
      &  \mat{I}_4 & \mat{0}_4 & -\mat{I}_4 &\\
      & & \ddots   &  \ddots   & \ddots 
    \end{array}
  \right],
\end{gather*}
where in the harmonic case
\begin{gather*}
  \skew3\widehat{\mat{B}} = \left[
    \begin{array}{cccc}
      0 & 2 & 0 &	0 \\
      0 & 0 & 0 & 0 \\
      0 & 0 & 0 &	2 \\
      0 & 0 & 0 & 0 
    \end{array}
  \right]
  \qquad \text{and} \qquad 	
  \skew3\widetilde{\mat{B}} = \mat{0}_4 ,
\end{gather*}
whereas in the subharmonic case
\begin{gather*}
  \skew3\widehat{\mat{B}} = \left[
    \begin{array}{rrrr}
      1 &  0 & 0 &  0 \\
      0 & -1 & 0 &  0 \\
      0 &  0 & 1 &  0 \\
      0 &  0 & 0 & -1 \\
    \end{array}
  \right] 
  \qquad \text{and} \qquad 
  \skew3\widetilde{\mat{B}} = -\mat{I}_4.
\end{gather*}

\begin{figure}
  \centering
  \includegraphics{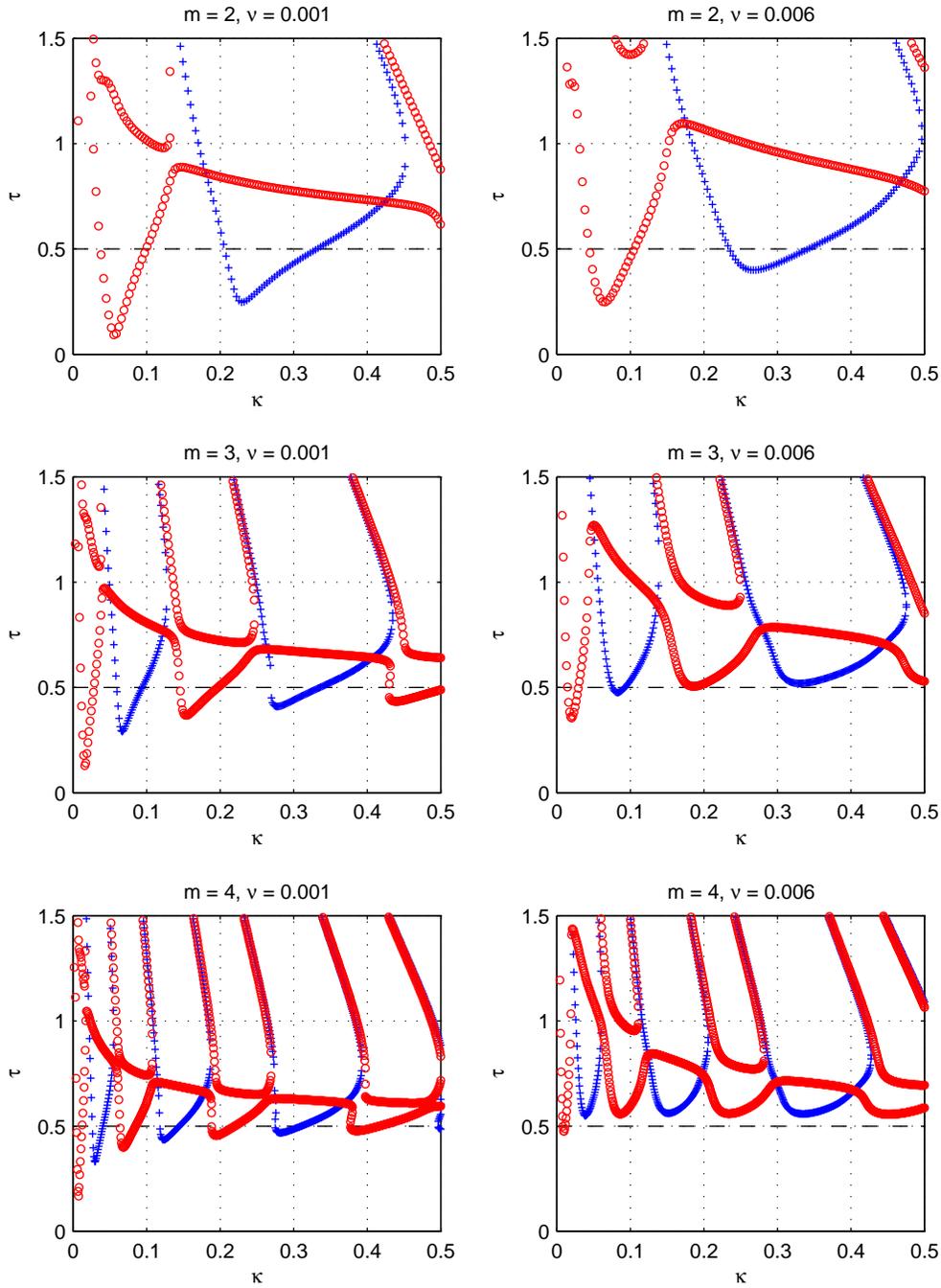}
  \caption{Ince-Strutt diagrams for the original viscous problem with
    dimensionless viscosities $\nu=10^{-3}$ (left) and
    $\nu=6\times10^{-3}$ (right).  Stability contours are shown for
    spherical harmonics $m=2$ (top), 3 (middle) and 4 (bottom).
    Subharmonic solutions are denoted with {\color{Red}$\bigcirc$} and
    harmonic solutions with {\color{Blue}$+$}.}
  \label{fig:contours-viscous}
\end{figure}

To illustrate the stability of the viscous problem, we solve the
eigenvalue equations for two values of the dimensionless viscosity, $\nu
= 10^{-3}$ and $6\times10^{-3}$, and for spherical harmonics numbered
$m=2, 3, 4$.  In the numerical calculations, we use a truncation size of
$N=80$ so that all trailing coefficients of $\{X_n^r, X_n^\Psi\}$ for
$N>80$ are no larger than $10^{-9}$ and so can be easily neglected. The
corresponding Ince-Strutt diagrams are shown in
figure~\ref{fig:contours-viscous} where again we observe a clearly
defined sequence of alternating harmonic and subharmonic fingers of
instability in parameter space.  These results reinforce one of the
defining characteristics of parametric resonance, namely that such
systems can experience instabilities leading to \emph{unbounded growth}
even in the presence of damping.

There are few key comparisons that can be drawn with the inviscid
results from section~\ref{sec:inviscid}.  First of all, the stability
fingers do not touch the $\kappa$-axis as they did in the inviscid case,
but instead are shifted vertically upwards.  As a result, there is a
minimum forcing amplitude required to initiate resonance for any given
value of $\kappa$.  For the smaller value of viscosity $\nu=10^{-3}$ the
fingers appear most similar to the inviscid case, while for larger $\nu$
the fingers deform upwards away from the $\kappa$-axis and shift outward
to the right.  Indeed, for large enough values of either viscosity or
spherical harmonic $m$ the fingers can lift entirely above the line
$\tau=\textstyle\half$ so that resonant behaviour is no longer possible.
This should be contrasted with the inviscid case where resonances exist
for any value of $m$.

The second distinction from the inviscid results is that within the
non-physical region $\textstyle \tau>\half$, an additional
subharmonic solution appears as a curve of circular points that cuts
across the finger-shaped contours.  These unstable modes occur due to
the periodic modulation in the tangential stress across the membrane
(\ref{eq:duPhi-jump}) and thus are not observed in the inviscid case.
However, all of these modes are restricted to the non-physical region
$\textstyle \tau > \half$ and so they can be considered as
spurious and safely ignored.

\section{Numerical simulations}
\label{sec:numerical}

Our next aim is to verify the existence of parametric instabilities for
an internally forced spherical membrane using numerical simulations of
the full governing equations~\eqref{eq:full-equations}.  We use a
parallel implementation of the immersed boundary algorithm of
\citet{Wiens2014} and \citet{WiensStockie} that utilises a
pseudo-compressible Navier-Stokes solver having particular advantages in
terms of parallel speed-up on distributed clusters.  The algorithm
exploits a rectangular fluid domain with periodic boundary
conditions, but we found that a cubic computational domain with side
length $6R$ is sufficiently large to avoid significant interference from
the adjacent periodic shells.  The elastic shell is
discretised using a triangulated mesh generated with the Matlab code
\texttt{distmesh} \citep{Persson2004}, wherein each vertex is an IB node
and each edge in the triangulation is a force-generating spring link
that joins adjacent nodes.  The elastic force generated by the deformed
membrane is then simply the sum of all spring forces arising from this
network of stretched springs.  The membrane is given an initial
configuration
\begin{gather*}
  \X(\xi,\eta,0) = R \big(1 + \epsilon
  Y_{m,k}^c(\xi,\eta)\big)\,\hat{r}, 
\end{gather*}
corresponding to a chosen scalar spherical harmonic of degree $m$ and order
$k$ with perturbation amplitude
\begin{gather*}
  \epsilon = \frac{0.05}{ \displaystyle \max_{\xi,\eta} \left|
      Y_{m,k}^c(\xi,\eta)\right|}\, .
\end{gather*}

\begin{figure}
  \centering
  \includegraphics{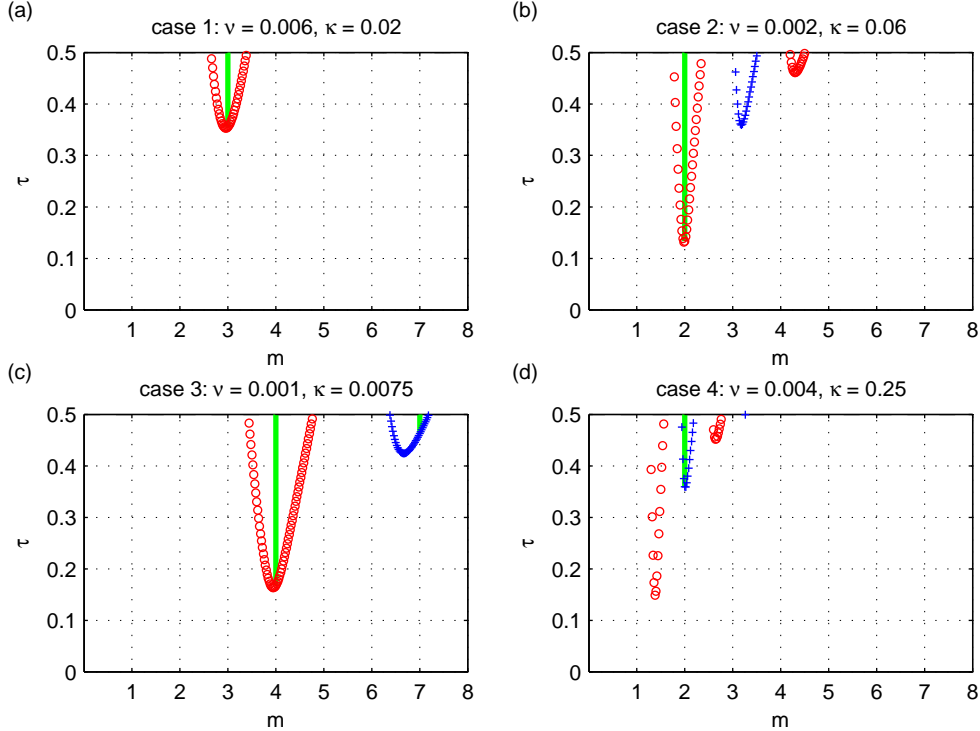}
  \caption{Ince-Strutt plots showing stability contours for each test
    case~1--4. Parameters that give rise to observable instabilities
    (i.e., corresponding to integer values of $m$) are highlighted by a
    vertical green line.}
  \label{fig:contours-sims}
\end{figure}

We then performed numerical simulations for four different sets of
parameters listed in table~\ref{tab:params}, which we refer to as cases
1--4.  This table lists the physical parameters used in the simulations
($\rho$, $\mu$, $R$, $\omega$, $\sigma$) as well as the corresponding
dimensionless parameters appearing in our analytical results ($\nu$,
$\kappa$).  The corresponding stability contours for each case 1--4 are
shown in figure~\ref{fig:contours-sims}, this time in terms of plots of
$\tau$ versus $m$ holding the values of $\nu$ and $\kappa$ fixed.  This
alternate view of the stability regions allows us to identify the
unstable modes that correspond to physical oscillations, since only
modes with integer values of $m$ are actually observable.
\begin{table}
  \begin{center}
    \begin{tabular}{cccccccccc}
      Case &\;\;& $\nu$ & $\kappa$ &\;\;& 
      $\rho$ ($\mbox{g/cm}^3$) &
      $R$ (cm) & $\omega$ ($1/\mbox{s}$) & $\mu$ ($\mbox{g/cm\,s}$) & 
      $\sigma$ ($\mbox{g/s}^2$) \\[3pt] 
      1 && 0.006 & 0.02  && 1 & 1   & 1    & 0.006 & 0.02 \\
      2 && 0.002 & 0.06  && 1 & 0.5 & 20   & 0.01  & 3    \\
      3 && 0.001 & 0.0075&& 1 & 0.5 & 20   & 0.005 & 0.375\\
      4 && 0.004 & 0.25  && 1 & 10  & 0.05 & 0.02  & 0.625\\
    \end{tabular}
    \caption{Dimensionless and physical parameters for the four
      test cases.}
    \label{tab:params}
  \end{center}
\end{table}

\begin{figure}
  \centering
  \includegraphics{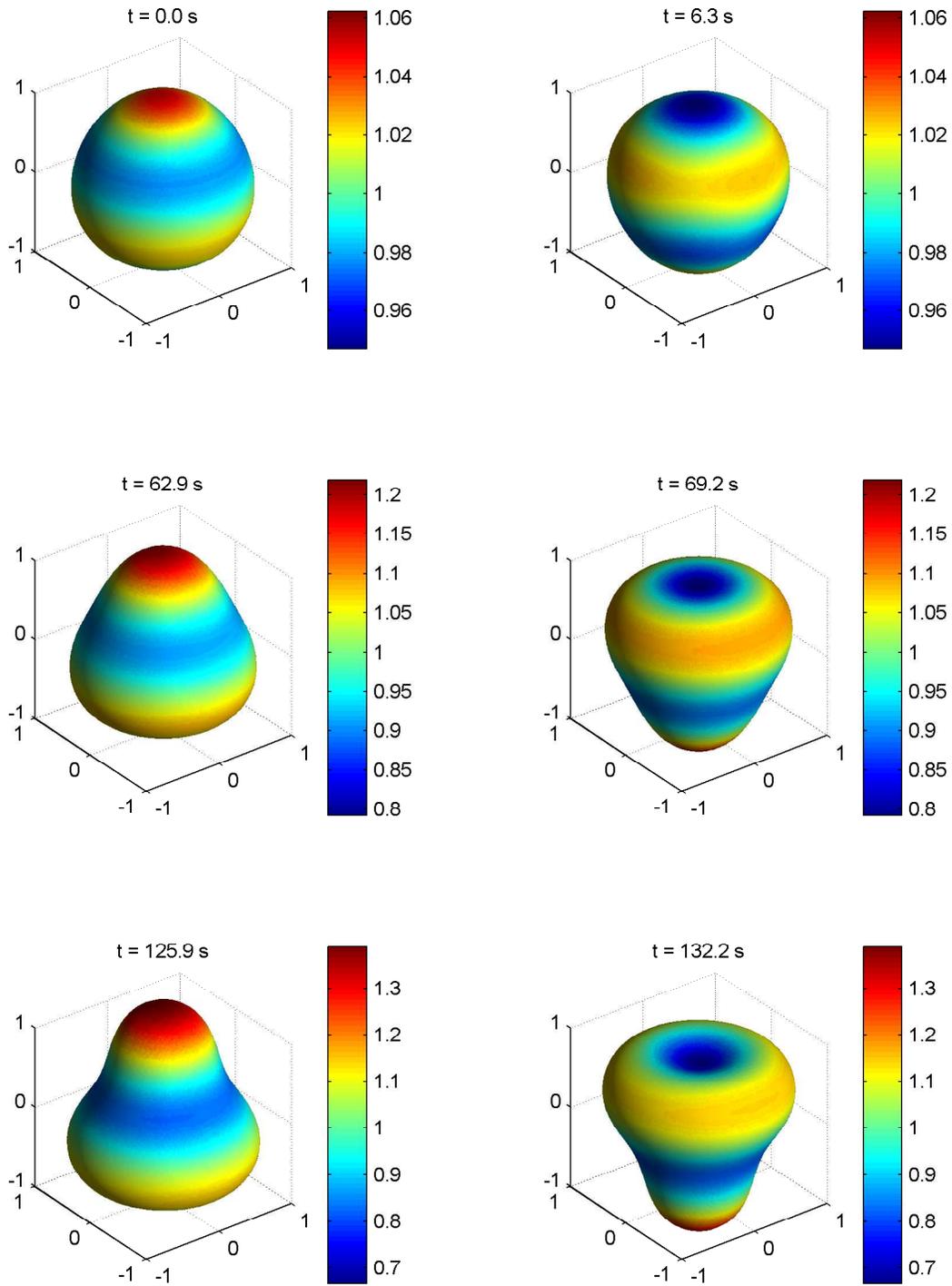}
  \caption{Several snapshots from the case~1 simulation where the
    membrane was initialised with a $(3,0)$-mode and the forcing
    amplitude is set to $\tau = 0.45$.}
  \label{fig:sim-snapshots}
\end{figure}

\begin{figure}
  \centering
  \includegraphics{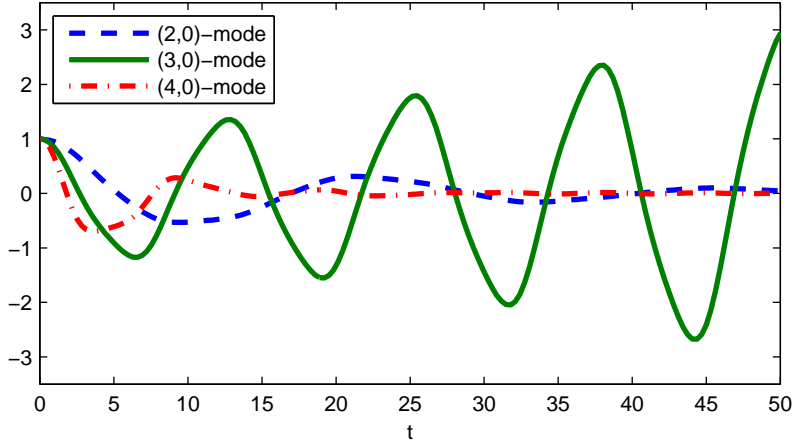}
  \caption{Radial amplitude projection $\widehat{X}^r$ for case~1, with
    the spherical shell perturbed by three different $(m,k)$-modes.  The
    results are all rescaled to start at $\widehat{X}^r=1$.}
  \label{fig:sim-compare-case1}
\end{figure}

Numerical simulations are performed by initialising the membrane with
various ($m$,$k$)-modes lying inside and outside the highlighted
unstable fingers so that direct comparisons can be drawn with our
analytical results.  Figure~\ref{fig:sim-snapshots} depicts several
snapshots of the numerical solution for case~1, where the membrane was
perturbed by a $(3,0)$-mode. The parametric forcing amplitude was
set to $\tau = 0.45$ which is well within the stability finger in 
figure~\ref{fig:contours-sims}a. Over time, the small initial perturbation
grows and oscillates with a frequency equal to one-half that of the
forcing frequency, as expected from the linear analysis.  To help
visualise the growth of this mode over time, we compute the projection
\begin{gather*}
  \widehat{X}^r(t) = \iints \X \bcdot \vshY^c
  \sin\eta\; \d\xi \; \d\eta  
\end{gather*}
at each time.  This integral is computed by first interpolating the IB
mesh values of $\X$ onto a regular $(\xi,\eta)$ grid and then
approximating the integral numerically using a Fast Fourier Transform in
$\xi$ and Gauss-Legendre quadrature in $\eta$.
Figure~\ref{fig:sim-compare-case1} depicts the evolution of
$\widehat{X}^r(t)$ for the $(3,0)$-mode, from which it is easy to see
the expected period-doubling response to a waveform with period $4\upi$.
To illustrate that the membrane instability depends sensitively on the
choice of mode, figure~\ref{fig:sim-compare-case1} also shows
simulations that were perturbed with $(2,0)$ and $(4,0)$-modes; neither
of these two other modes exhibits any evidence of instability, which is
also predicted by the stability plots in figure~\ref{fig:contours-sims}a.

\begin{figure}
  \centering
  \includegraphics{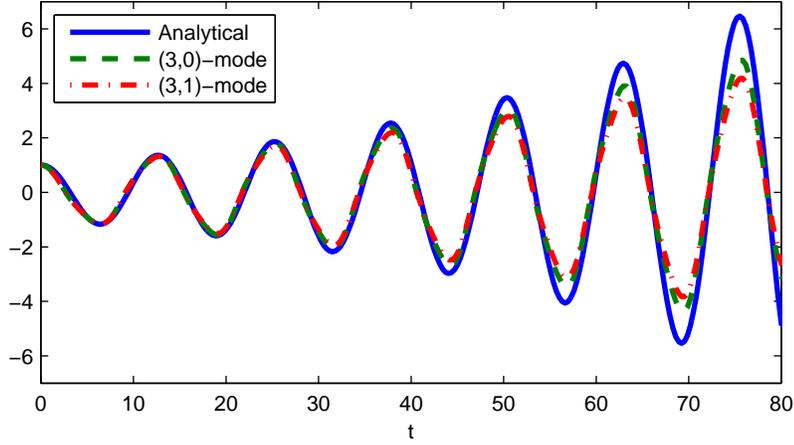}
  \caption{Radial amplitude projection $\widehat{X}^r$ in case 1,
    showing a spherical shell perturbed by a $(3,0)$-mode (dashed) and a
    $(3,1)$-mode (dash-dot).  The exact result from the linear analysis
    is shown as a solid line.  All curves are rescaled to start at
    $\widehat{X}^r=1$.}
  \label{fig:sim-compare-m3}
\end{figure}

One conclusion from our analysis is that for each $m$, the
stability of the linearised dynamics does not depend on the order $k$ of
the spherical harmonic.  This result is investigated in
figure~\ref{fig:sim-compare-m3}, where we display the projected radial
amplitude $\widehat{X}^r$ for two simulations using the case~1
parameters and two initial membrane shapes corresponding to modes
$(3,0)$ and $(3,1)$.  The analytical solution is provided for
comparison, and we observe that the behaviour of all three solution
curves is nearly indistinguishable at early times.  However, as time
goes on the perturbations grow and non-linear affects come into play in
the simulations, so that the growth rates eventually deviate from the
linear analysis.

\begin{figure}
  \centering
  \includegraphics{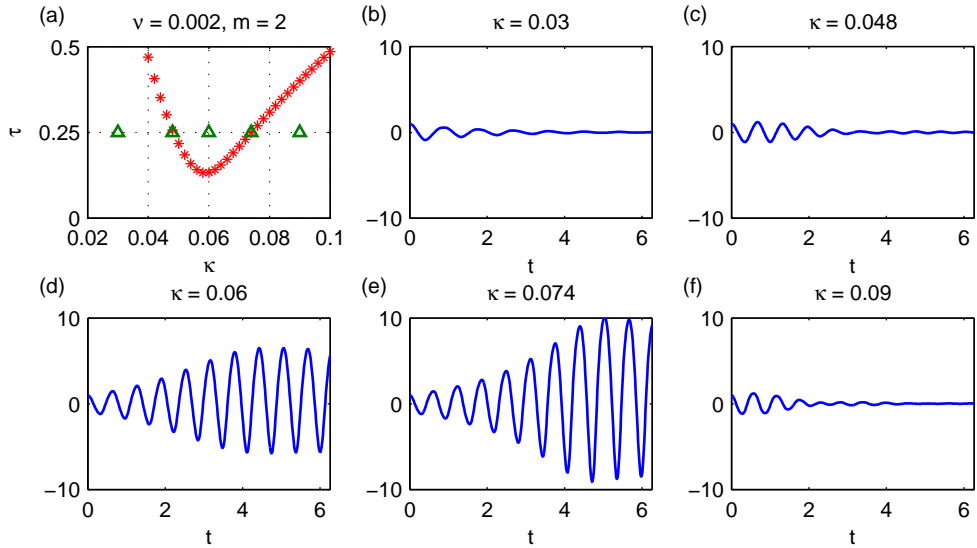}
  \caption{(a) Ince-Strutt diagram for $m=2$ and $\nu = 0.002$ but with
    varying $\kappa$. The parameters used are denoted by
    {\color{OliveGreen}$\triangle$}, with the centre point corresponding
    to case~2.  (b)--(f) Radial projection from simulations for a range
    of $\kappa$ values.  The membrane was initialised with a
    $(2,0)$-mode and all curves are rescaled to start at
    $\widehat{X}^r=1$.}
  \label{fig:sim-compare-kappa}
\end{figure}

As another illustration of the dependence of stability on system
parameters, we perform a series of simulations to investigate the
``sharpness'' of the stability fingers.  Using
parameters from case~2, we fix $\tau = 0.25$ and then investigate the
behaviour of the numerical solution as $\kappa$ is varied.  Based on
figure~\ref{fig:contours-sims}b, we expect case~2 to be unstable if the
membrane configuration is an $m=2$ mode, but if parameters are changed
sufficiently then the stability fingers can shift enough that the 2-mode
stabilises.  Figure~\ref{fig:sim-compare-kappa}a shows the Ince-Strutt
diagram as a plot of $\tau$ versus $\kappa$ for $\nu=0.002$ and $m=2$.
The parameter values used in this series of tests are denoted by
{\color{OliveGreen}$\triangle$} in figure~\ref{fig:sim-compare-kappa}a,
with the centre point located in the middle of the subharmonic finger
corresponding to case~2, and the remaining parameters lying either on
the border of the stability region ($\kappa\approx 0.048$ and 0.074) or
outside.  Figures~\ref{fig:sim-compare-kappa}b--f depict the radial
projection $\widehat{X}^r$ for each of the five simulations from which
we can clearly see that as $\kappa$ increases the solutions transition
from stable to unstable and then back to stable again.  The computed
stability boundaries do not correspond exactly with the analytical
results, particularly for the upper stability limit $\kappa\approx
0.074$; however, the match is still reasonable considering that the
analysis is linear.

\begin{figure}
  \centering
  \includegraphics{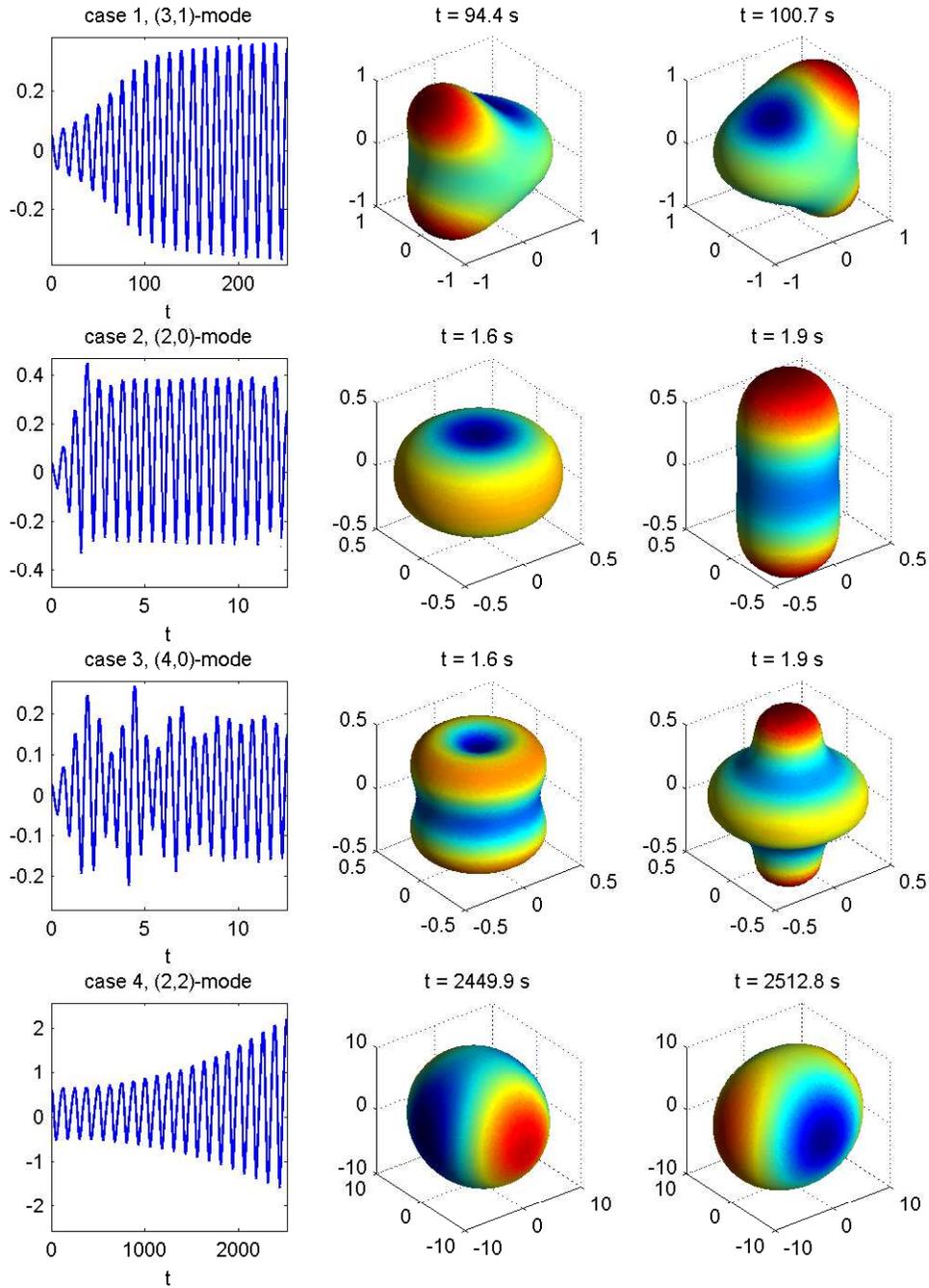}
  \caption{Simulation results for all four cases where a spherical shell
    perturbed by a mode that is expected to be unstable, with $\tau =
    0.45$. The left column shows the radial projection $\widehat{X}^r$,
    while snapshots from the corresponding simulations are shown on the
    right.}
  \label{fig:sim-all}
\end{figure}

Finally, we show in figure~\ref{fig:sim-all} plots of the radial
amplitude projection in cases 1--4, perturbing the spherical shell in
each case with a mode that we know from the analysis to be unstable.
The forcing amplitude is set to $\tau = 0.45$ in all four cases.
Snapshots of the membrane evolution are also given for each simulation
to illustrate the growth of the given modes. All simulations exhibit the
expected unstable growth in solution amplitude, although we stress that
the numerical results do not lead to unbounded growth (or blow-up) in
the amplitude as suggested by the linear analysis.  We attribute this
discrepancy in behaviour to nonlinear effects that become important
later in the simulation and limit the solution growth when the amplitude
of oscillations become large enough.  We also note that the correct
frequency response is observed in all four tests, with cases 1--3
exhibiting the expected period-doubling subharmonic response, while 
case~4 oscillates with a harmonic response.

\section{Application to the heart and other biofluid systems}
\label{sec:biofluids}

Our work on parametric forcing of immersed elastic shells was originally
motivated by the study of actively-beating heart muscle fibres that
interact with surrounding blood and tissue.  Heart muscle contractions
are actually initiated by complex waves of electrical signals that
propagate through the heart wall, which should be contrasted with the
spatially-uniform coordinated contractions analysed in this paper.
Furthermore, the heart chambers have an irregular shape and a thick
wall that differ significantly from a spherical shell with zero thickness.
Nonetheless, it is natural to ask whether our analysis of spherical
immersed elastic shells with a periodic internal forcing could still
yield any useful insights into the nature of the complex fluid-structure
interaction between a beating heart and the surrounding fluid.  

To this end, we consider an immersed spherical shell with parameters
that correspond to the human heart under two conditions: first, a normal
healthy heart; and second, an abnormal heart undergoing a much faster
heartbeat.  We then ask whether our analysis gives rise to resonant
behaviour in either case for physiologically relevant heart beat
frequencies.  There are a wide range of abnormal heart rhythms
classified under the heading of \emph{supraventricular tachycardia} or
{SVT}~\citep{Link2012,PhillipsFeeney1973}, corresponding to a heart
rhythm that is either irregular or abnormally rapid and occurs in the
heart's upper chambers (called the left and right atria).  In contrast
with ventricular tachycardias, which are much more dangerous, many SVTs
are non-life-threatening and can persist for long periods of time.
Therefore, we will focus on SVTs (and the atria) where fluid dynamic
instabilities are more likely to have the time to develop.

We next discuss the choice of parameters that is appropriate for 
applying our IB model to study FSI in the heart.  The resting heart rate
for a healthy person ranges from 60 to 100 beats per minute (bpm) and
both atria and ventricles beat in synchrony.  
In contrast, a heart characterised by SVT can exhibit two separate beats
in the atria and ventricles, and can have an atrial rhythm that lies
anywhere between 100 to 600~bpm.
A clinical study by \citet{Wang2011} surveyed 322 patients suffering
from atrial fibrillation (one sub-class of SVT) and obtained
measurements of atrial wall stiffness $\sigma$ varying between $1\times
10^3$ and $2\times 10^4\;\units{dyn/cm}$.  We have found no evidence to
suggest that the stiffness varies significantly between hearts with
normal and abnormal rhythms, and so we use the same range of $\sigma$
for all cases.
Although hearts suffering from conditions such as atrial fibrillation
are often characterised by an increased size \citep{Wang2011}, we elect
to use a single representative value of the radius $R=2.0\;\units{cm}$
for an atrium in both normal and diseased hearts.  In terms of the fluid
properties, blood has density similar to water with
$\rho=1\;\units{g/cm^3}$ but has a significantly higher dynamic
viscosity of $\mu=0.04\;\units{g/cm\,s}$.  We then choose a
representative SVT rhythm with frequency $\omega=400$\;bpm, which
translates into a dimensionless viscosity parameter $\nu=2.39 \times
10^{-4}$.

\begin{figure}
  \centering
  \includegraphics{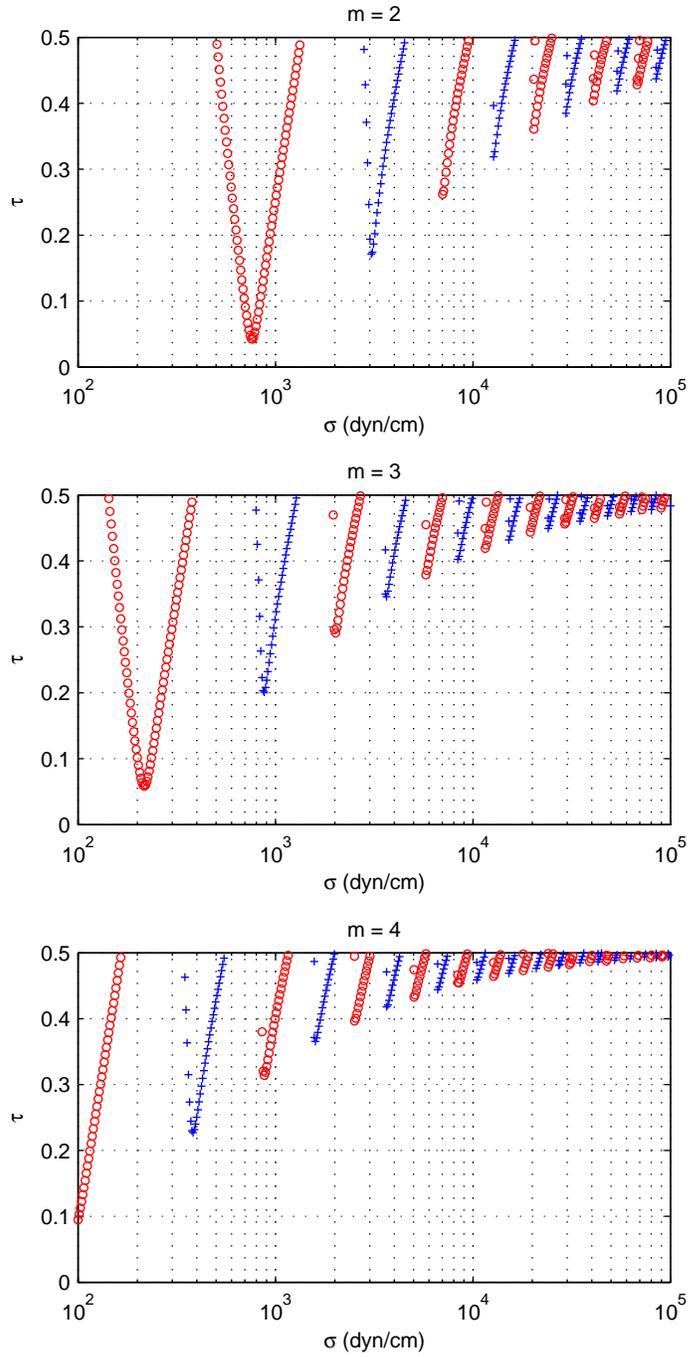}
  \caption{Stability contours for physical parameters corresponding to a
    human heart undergoing an abnormal heart rhythm ($\nu = 2.39 \times
    10^{-4}$) for modes $m = 2$ (top), 3 (middle) and 4 (bottom).}
  \label{fig:contours-heart}
\end{figure}

Substituting the parameter values and ranges just described into our
analytical results from section~\ref{sec:viscous} for the first three
modes numbered $m=2,3,4$, we obtain the Ince-Strutt plot in
figure~\ref{fig:contours-heart}.  The stability fingers are depicted
with the elastic stiffness parameter $\sigma$ plotted along the
horizontal axis, and the results show that parametric instabilities can
arise for most values of $\sigma$ under consideration.  Because these
resonant instabilities occur over such a wide range of stiffness values
covered by the measurements by \citet{Wang2011} from the left atrium, it
is reasonable to hypothesise that it may be possible for FSI-driven
parametric instability to influence the dynamics of the beating heart.

\begin{figure}
  \centering
  \includegraphics{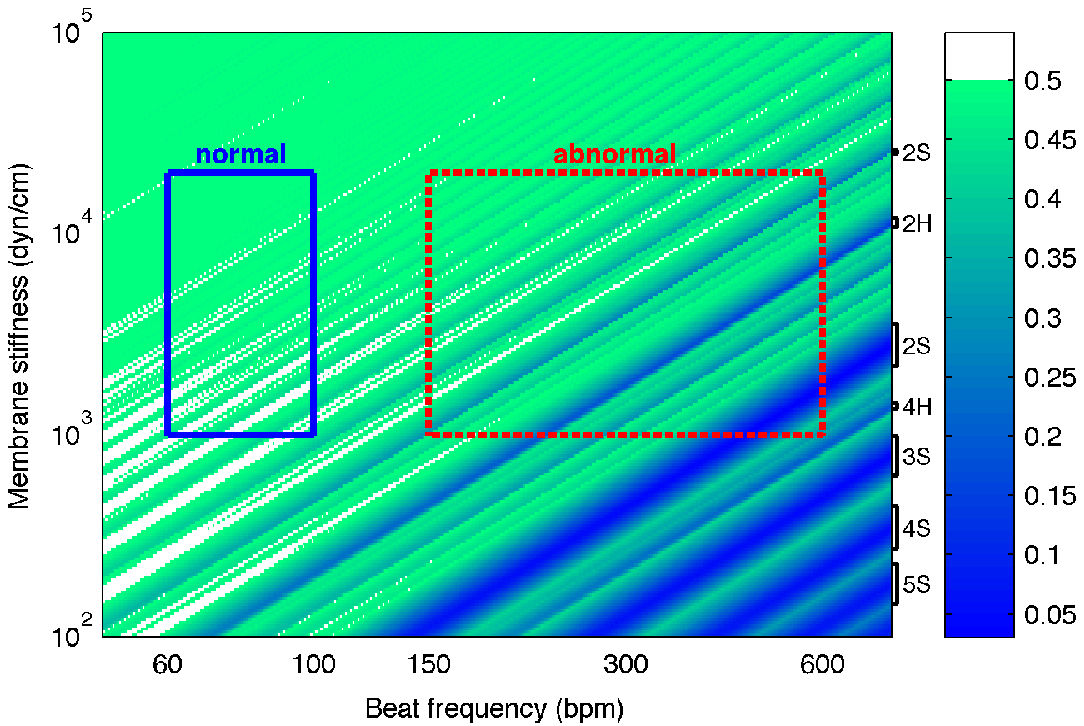}
  \caption{Minimum value of $\tau$ required for parametric resonance
    across modes 2--6. White areas represent parameters for which no
    physical instability exists. Each dark (blue) band is labelled by its
    corresponding unstable $m$-mode and response frequency 
    (`H' for harmonic or `S' for subharmonic). The boxes indicate parameter ranges
    for a healthy heart (solid, blue) and for a heart with SVT (dashed,
    red).}
  \label{fig:min-tau}
\end{figure}

We now delve further into the finger plots in
figure~\ref{fig:contours-heart} and remark that experiments suggest the
heart muscle is seldom (if ever) completely slack; therefore, we expect
that the forcing amplitude parameter $\tau$ lies significantly below the
threshold value of $\textstyle\half$.  Indeed, an estimate of
$\tau\approx 0.3$ can be extracted from measurements of pressure in the
left atrium taken by \citet[Fig.~1]{BraunwaldEtal1961}.
Figure~\ref{fig:min-tau} provides an alternate view of the dependence of
resonant instabilities on the parameters by depicting the minimum $\tau$
giving rise to resonance for as a function of elastic stiffness and beat
frequency (using modes in the range $m=2$--6).  We are especially
interested in the dark (blue) bands that correspond to smaller values of
$\tau$ and hence more prominent instabilities.  Taking a value of
$\tau=0.3$, our analysis predicts ``valleys'' of instability
corresponding to discrete ranges of the parameters $\omega$ and
$\sigma$.  For fixed $\sigma$, we observe that at low frequency these
valleys are very narrow and steep, while as the forcing frequency
increases the width of the unstable bands likewise increases.  In
particular, if we consider an intermediate value of $\sigma$ for
``normal'' heart beating in the range of 60--100~\units{bpm}, then the
unstable bands are relatively small and so resonances would seem to be
less likely.  On the other hand, if the frequency is increased to
300--600~\units{bpm} then we begin to encounter wider valleys that
suggest instabilities for a smaller value of $\tau$.

In summary, we have found that resonant instabilities are possible for a
wide range of parameters corresponding to both normal and abnormal
hearts; furthermore, for higher frequencies corresponding to SVT,
instabilities are not only more likely to occur but also persist over
wider ranges of parameter space.  As suggestive as these results are, we
refrain from making any specific claims or predictions regarding
resonance in the actual beating heart since we have made so many
simplifications and assumptions here: the applied periodic forcing is
oversimplified, nonlinearities are neglected, a spherical shell is a far
cry from an atrium, and parameter values are still quite uncertain.
Nonetheless, the fact that our analysis predicts resonances for such a
wide range of physiologically relevant parameters is compelling enough
to suggest that this problem merit further investigation.  Moreover, it
should be possible to test for the presence of isolated parametric
instabilities in carefully designed experiments, and our parameter study
above provides guidance in what ranges of parameters are most worthy of
investigation.

In addition to the heart, there are many other complex bio-fluid systems
that undergo similar periodic internal forcing and for which our
analytical results could potentially be applied.  Parametric forcing in
spherical membranes occurs in the context of dielectric characterisation
and manipulation of biological cells (namely, erythrocytes and
lymphocytes) by way of an applied electric field~\citep{Zehe2004b}.
Another problem with spherical geometry was considered by
\citet{Cottet2006a}, who were motivated by the study of cell membrane
protrusions appearing in the process of cell locomotion.  A much simpler
geometry in which we have already applied a similar Floquet analysis is
an IB model for the basilar membrane in the inner
ear~\citep{KoStockie2014}.  This problem involved a flat membrane for
which the Floquet expansion takes a correspondingly simpler form in
terms of trigonometric eigenfunctions.  We showed that parametric
resonances may play a significant role in the astounding ability of the
mammalian ear to sense and process sound.

Another common geometry that is intermediate in complexity between the
flat membrane and spherical shell is a cylindrical tube, which plays a
central role in many biological structures (such as arteries, or bronchi
in the lungs) as well as in engineering systems (involving compliant
pipes or tubes).  Indeed, there are many examples of FSI systems
involving flow through a compliant cylindrical tube wherein some form of
periodic forcing is known to induce resonance, including blood flow
through artery-organ systems~\citep{WangEtal1991}, 
air flow-induced vibrations of the bronchi~\citep{Grotberg1994}, and the
resonant impedance pump~\citep{AvrahamiGharib2008, LoumesEtal2008}
whose design was incidentally inspired by observations of resonant
pumping in the embryonic heart.  An exciting possible avenue for future
work is to extend our Floquet-type analysis to a cylindrical
shell, which could then provide new insight into parametric resonance
instabilities arising in these other FSI systems.  One obvious
advantage to studying this cylindrical geometry is that we can avoid the
complexity of vector spherical harmonics and instead employ a simpler
Floquet series solution consisting of Fourier-Bessel eigenfunctions,
similar to the 2D radial solution obtained for a circular membrane in
\citet{Cortez2004}.

\section{Conclusions}
\label{sec:conclusions}

In this paper, we have demonstrated the existence of parametric
instabilities in a spherical elastic shell immersed in an incompressible
viscous fluid, wherein the motion is driven by periodic contractions of
the shell.  A mathematical model was derived using an immersed boundary
framework that captures the full two-way interaction between the elastic
material and the surrounding fluid.  A Floquet analysis of the
linearised governing equations is performed using an expansion in terms
of vector spherical harmonics.  We obtained results regarding the
stability of the internally forced system with and without viscosity,
and showed with the aid of Ince-Strutt diagrams that fluid-mechanical
resonance exists regardless of whether viscous damping is present.
Numerical simulations of the full IB model were performed that confirm
the presence of these parametric resonances.  In parallel with this
work, we have initiated an experimental study of rubber water balloons
immersed in water~\citep{KoTseStockie} in which we measure the natural
modes of oscillation of these immersed (nearly-spherical) membranes and
use the results to further validate our analytical predictions.

Because our original motivation for considering this problem derived
from the study of periodic contractions driving blood flow in the heart,
we also discussed the relevance of our stability analysis to cardiac
fluid dynamics.  Indeed, our analysis suggests that periodic resonances
can occur in an idealised spherical shell geometry for physical
parameters corresponding to the heart, provides possible parameter
ranges to investigate in an experimental study that could test for
resonant solutions.  These results are very preliminary and much more
work needs to be done to determine whether fluid-structure driven
resonances can actually play a role in cardiac flows.

One major step in bringing our results closer to the actual heart is to
generalise our time-periodic (but spatially uniform) driving force to
include the effect of spiral waves of contraction that are initiated
through electrical signals propagating in the heart wall.  Including
such spatiotemporal variations in the driving force would naturally
couple together the radial and angular solution components.  As a
result, we would then lose the special advantage we gained from our
choice of vector spherical harmonics that led to a fortunate
mode-decoupling in the interfacial jump conditions.  Generalising the
analysis to handle this fully coupled problem would require a
considerable effort, but could lead to significant new insights into a
more realistic model of FSI in the beating heart.
\\

Thanks go to Jeffrey Wiens for permitting us to use his immersed
boundary code, as well as for providing assistance with the
modifications required to perform the simulations in this paper.  
Financial support for
this work came from a Discovery Grant and a Discovery Accelerator Award
from the Natural Sciences and Engineering Research Council of Canada.



\bibliographystyle{jfm}
\bibliography{resonance3DIB}

\end{document}